\documentclass{aa}  

\usepackage{graphicx}
\usepackage{txfonts}
\usepackage{epsfig}
\usepackage{longtable}
\usepackage{lscape}
\usepackage{amssymb} 
\usepackage{hyperref}
\usepackage{refcount}

\newcommand{\sw}{$Swift$}
\def \inte {INTEGRAL}
\def \rxte {RXTE}
\def \sax {{\em BeppoSAX}}
\def \suzaku {{\em Suzaku}}
\def \sw {{\em Swift}}
\def \xmm {{\em XMM--Newton}}

\def \hcm {\hbox {\ifmmode $ atom cm$^{-2}\else atom cm$^{-2}$\fi}}

\def \arcsec {\hbox{$^{\prime\prime}$}}

\def \ATel {ATel}
\def \iaucirc {IAU Circ.}
\def \apj {ApJ}
\def \apjl {ApJL}
\def \apjs {ApJS}\def \aap {A\&A}

\def \baas {BAAS}
\def \pasj {PASJ}
 
\def \mnras {MNRAS}
\def \ssr {Space Science Reviews}

\begin{document} 

\title{The 100-month \emph{Swift} catalogue of supergiant fast X--ray transients\thanks{ \href{http://www.ifc.inaf.it/sfxt/}{Project web page: http://www.ifc.inaf.it/sfxt/ .} }}
   \subtitle{I. BAT on-board and transient monitor flares\thanks{Table~\ref{sfxtcat:tab:catalogue} is 
only available in electronic form at the CDS via anonymous ftp to 
                                    {\tt cdsarc.u-strasbg.fr} (130.79.128.5) or via 
                                    \href{http://cdsweb.u-strasbg.fr/cgi-bin/qcat?J/A+A/XXX/XXX}{{\tt http://cdsweb.u-strasbg.fr/cgi-bin/qcat?J/A+A/XXX/XXX}
                                            } 
             } 
} 

\author{P.\ Romano\inst{1}
          \and 
          H.A.\ Krimm\inst{2,3}
         \and
           D.M.\ Palmer\inst{4}
        \and
        L.\ Ducci \inst{5} 
        \and
         P. Esposito\inst{6}  
         \and
         S. Vercellone\inst{1} 
          \and
         P.A. Evans\inst{7} 
           \and
          C.~Guidorzi\inst{8}
        \and 
          V.~Mangano\inst{9}
         \and    
          J.A.~Kennea\inst{9}
          \and
          S.D.~Barthelmy\inst{2}
          \and
          D.N.~Burrows\inst{9}
          \and 
          N.~Gehrels\inst{2}
          }
   \institute{INAF, Istituto di Astrofisica Spaziale e Fisica Cosmica - Palermo,
              Via U.\ La Malfa 153, I-90146 Palermo, Italy \\
              \email{romano@ifc.inaf.it} 
              \and
              NASA/Goddard Space Flight Center, Greenbelt, MD 20771, USA
              \and
              Universities Space Research Association, Columbia, MD, USA 
              \and
              Los Alamos National Laboratory, B244, Los Alamos, NM 87545, USA
               \and 
              Institut f\"ur Astronomie und Astrophysik, Eberhard Karls Universit\"at, 
              Sand 1, 72076 T\"ubingen, Germany
               \and
             INAF, Istituto di Astrofisica Spaziale e Fisica Cosmica - Milano, 
              Via E.\ Bassini 15,   I-20133 Milano,  Italy 
             \and
              University of Leicester, X-ray and Observational Astronomy Group, 
              Department of Physics \& Astronomy, University Road, \\
              Leicester, LE1 7RH
              \and
              Department of Physics and Earth Sciences, University of Ferrara,
              Via Saragat 1, I-44122, Ferrara, Italy 
              \and
              Department of Astronomy and Astrophysics, Pennsylvania State 
              University, University Park, PA 16802, USA
              }
 \date{Received August 21, 2013; accepted November 27, 2013}

\abstract{ 
Supergiant fast X--ray transients (SFXTs) are high mass X--ray binaries (HMXBs) 
that are defined by their hard X--ray flaring behaviour. During these flares 
they reach peak luminosities of 10$^{36}$--10$^{37}$~erg~s$^{-1}$ for 
a few hours (in the hard X--ray) which are much shorter timescales 
than those characterizing Be/X--ray binaries. 
}{ 
We investigate the characteristics of bright flares (detections in excess of $5\sigma$) 
for a sample of SFXTs and their relation to the orbital phase. 
}{ 
We have retrieved all {\em Swift}/BAT Transient Monitor light curves and 
collected all detections in excess of $5\sigma$ from both 
daily- and orbital-averaged light curves in the time range of 
2005 February 12 to 2013 May 31 (MJD~53413--56443). 
We also considered all on-board detections as recorded in the same 
time span  and selected those in excess of $5\sigma$ and within 4 arcmin of each source in our sample. 
}{ 
We present a catalogue of over a thousand BAT flares from 11 SFXTs, 
down to 15--150\,keV fluxes of  
$\sim6\times10^{-10}$ erg cm$^{-2}$ s$^{-1}$  (daily timescale) and  
$\sim 1.5\times10^{-9}$ erg cm$^{-2}$ s$^{-1}$ (orbital timescale, averaging $\sim 800$\,s); 
the great majority of these flares are unpublished. 
The catalogue spans 100 months. 
This population is characterized by short (a few hundred seconds) and relatively bright 
(in excess of 100\,mCrab, 15--50\,keV) events. In the hard X--ray, these flares last generally  
much less than a day.  
Clustering of hard X--ray flares can be used to indirectly measure the length of an outburst, 
even when the low-level emission is not detected. 
We construct the distributions of flares, of their significance 
(in terms of $\sigma$), and of their flux as a function of orbital phase  
to infer the properties of these binary systems.  
In particular, we observe a trend of clustering of flares at some phases as $P_{\rm orb}$ increases,  
which is consistent with a progression from tight circular or mildly eccentric orbits at short periods  
to wider and more eccentric orbits at longer orbital periods. 
Finally, we estimate the expected number of flares for a given source for our limiting flux and
provide the recipe for calculating them for the limiting flux of future hard X--ray observatories.  
}{%
} 
 
    
   \maketitle
%

	\section{Introduction\label{sfxtcat:intro}}

%
Supergiant fast X--ray transients (SFXTs) are a class of high mass
X--ray binaries (HMXBs) associated with OB supergiant stars
 brought to the foreground by \inte\ 
\citep[][]{Smith2004:fast_transients,Sguera2005,Negueruela2006:ESASP604}. 
They display sporadic X--ray outbursts characterized by bright flares 
with peak luminosities of 10$^{36}$--10$^{37}$~erg s$^{-1}$ that 
last for a few hours \citep[as observed by \inte; ][]{Sguera2005,Negueruela2006}, which are 
 significantly shorter than those of typical Be/X--ray binaries. 
In quiescence \citep[e.g.][]{zand2005,Bozzo2010:quiesc1739n08408},  
they have a luminosity of $\sim 10^{32}$~erg~s$^{-1}$;  therefore,  
one defining characteristic of SFXTs is the dynamic range of 3--5 orders of magnitude. 
During outburst their hard X--ray spectra resemble those of HMXBs 
that host accreting neutron stars (NS),  
with hard power laws below 10\,keV,  which are combined 
with high energy cut-offs at $\sim 15$--30~keV and 
sometimes strongly absorbed at soft energies \citep{Walter2006}. 
Therefore, it is tempting to assume that all SFXTs might host a neutron star, 
even if pulse periods have only been measured for just a few of them. 
The mechanism producing the outbursts is still subject to debate  
but it is probably related to either the properties of 
the wind from the supergiant companion 
\citep{zand2005,Walter2007,Negueruela2008,Sidoli2007} or the 
presence of gated mechanisms \citep[][]{Grebenev2007,Bozzo2008}. 
Given the typical timescales of SFXT activity, it is convenient to distinguish between 
a {\it flare}, which is a state of enhanced emission generally lasting for a few hours, 
and an {\it outburst}, which is composed of several flares and lasts for about a day or more
\citep[e.g.][]{Romano2007,Rampy2009:suzaku17544}.

\sw\ \citep{Gehrels2004} has been shedding new light on the phenomenon of SFXTs, 
thanks to its unique properties of automatic fast-slewing and broad-band energy coverage, 
that make it the only observatory which can detect outbursts from SFXTs 
from the very beginning and observe their evolution panchromatically. 
Indeed, the bright flares from SFXTs have been triggering 
the Burst Alert Telescope \citep[BAT, ][]{Barthelmy2005:BAT}
since early after launch. 
However, while the most spectacular evidence of SFXT activity comes from their outbursts, 
SFXTs are characterized by flares in all intensity states 
\citep[e.g.][]{Bozzo2010:quiesc1739n08408,Bodaghee2011:17391-3021}. 
Several ks-long bright flares were, for instance, observed with the X--ray Telescope \citep[XRT, ][]{Burrows2005:XRT} 
during the monitoring campaigns that were performed by taking advantage of \sw's scheduling 
flexibility, on 4 SFXTs during 2007--2009 
\citep{Sidoli2008:sfxts_paperI,Romano2009:sfxts_paperV,Romano2011:sfxts_paperVI}. 

As the BAT observes 88\% of the sky daily, on average,
it is ideally suited to detect flaring in hard X--ray astrophysical sources.  
Since 2005 February 12, the BAT Transient Monitor\footnote{
\href{http://swift.gsfc.nasa.gov/docs/swift/results/transients/}{http://swift.gsfc.nasa.gov/docs/swift/results/transients/. }}
\citep{Krimm2013:BATTM} has been providing 
near real-time light curves in the 15--50\,keV energy range of more than 900 sources
with a mean variance for one-day mosaics of 5.3\,mCrab. 
Several flares from SFXTs are regularly caught every year by the BAT Transient Monitor. 

Finally, BAT maintains a per-source flux threshold in its on-board catalogue.  
A source detection does not result in a Gamma-ray Coordinates Network 
(GCN)\footnote{\href{http://gcn.gsfc.nasa.gov}{http://gcn.gsfc.nasa.gov . }} Notice 
and possible follow-up observations 
unless it exceeds its threshold.  When BAT responds to a source, it automatically raises 
the threshold so that repeated bursts only cause a further response if they are 
significantly larger than earlier ones.  
The source threshold can be manually lowered to re-enable automatic observations 
on its next burst and such is the case for SFXTs. 

In this paper, we present for the first time the catalogue of all flares, 
the great majority of which are unpublished, 
registered by the BAT both on-board and on the ground by the Transient Monitor
between 2005 February 12 and  2013 May 31. 
The SFXT sample is defined in Sect.~\ref{sfxtcat:sample} along with a review of 
the most relevant literature. 
The methods for reduction and analysis and our results are presented in Sect.~\ref{sfxtcat:dataredu}. 
In a companion paper (P.\ Romano, in preparation), we present an in depth, broadband analysis of 
all the outbursts that triggered the BAT during the same time span.

\setcounter{table}{0} 
\begin{sidewaystable*}
 \tabcolsep 4pt   	
 \caption{Sample of confirmed SFXTs: Positions and optical counterparts.\label{sfxtcat:tab:datapositions} }
 \centering
 \begin{tabular}{lllrrrrcccc}
\hline \hline
\noalign{\smallskip} 
 Name              & RA        &Dec                                            &Error        & IR/Optical                          & Spectral &  Distance                         &   \multicolumn{4}{c}{References}  \\
                      & (J2000)  &(J2000)                                       &                  &   Counterpart                              & Type       & (kpc)                                & Discovery  & Position  & Sp. Type & Distance\\
 \hline  \noalign{\smallskip}
IGR~J08408$-$4503  &08 40 47.83  &$-$45 03 31.1   &1\farcs6       &HD~74194                                   &O8.5Ib(f)              &2.7                              &1          &2              &3            & 4      \\  
IGR~J11215$-$5952  &11 21 46.9    &$-$59 51 46.9   &1\farcs1       &HD~306414                                 &B0.7Ia                  &6.2                              &5          &6              &7            & 8      \\  
IGR~J16328$-$4726  &16 32 37.87  &$-$47 23 41.2   &1\farcs4       &2MASS~J16323791$-$4723409  & O8Iafpe              &3--10                         &9          &10           &11         &12  \\   
IGR~J16418$-$4532  &16 41 50.65  &$-$45 32 27.3  &1\farcs9        &2MASS~J16415078$-$4532253  &BN0.5Ia               &13                               &13        &14            &15          &15   \\  
IGR~J16465$-$4507  &16 46 35.5    &$-$45 07 04     &4\arcsec        &2MASS~J16463526$-$4507045  &B0.5Ib/O9.5Ia     &9.4/$9.5^{+14.1}_{-5.7}$ &16     &17           &18,15        &15,19 \\  
IGR~J16479$-$4514  &16 48 06.58  &$-$45 12 06.74 &0\farcs60     &2MASS~J16480656$-$4512068  &O8.5I                   &4.9                              &20        &21            &15          &15     \\ 
XTE~J1739$-$302	  &17 39 11.58   &$-$30 20 37.6  &0\farcs6       &2MASS~J17391155$-$3020380  &O8Iab(f)               &2.7                              &22,23   &24            &25           &15     \\  
IGR~J17544$-$2619  &17 54 25.284 &$-$26 19 52.62&0\farcs6       &2MASS~J17542527$-$2619526  &O9Ib                    &3.6                              &26        &27            &28          &15      \\ 
SAX~J1818.6$-$1703 &18 18 37.89  &$-$17 02 47.9   &0\farcs6      &2MASS~J18183790$-$1702479  &O9I-B1I                &$2.1\pm0.1$              &29	   &30             &31           & 32       \\  
AX~J1841.0$-$0536  &18 41 00.54   &$-$05 35 46.8  &0\farcs6       &2MASS~J18410043$-$0535465  &B1Ib                     &$3.2_{-1.5}^{+2.0}$,$6.9\pm1.7$  &33,34  &35 & 36   & 36,37 \\  
AX~J1845.0$-$0433  &18 45 01.58   &$-$04 33 57.4  &1\farcs4       &2MASS~J18450159$-$0433565  &O9.5I                   &3.6                              &38        &10           &39           &39   \\  
IGR~J18483$-$0311  &18 48 17.2     &$-$03 10 16.8  &0\farcs8       &2MASS~J18481720$-$0310168  &B0.5Ia/B0-B1Iab &$2.83\pm0.05$          &40       &41            &15,32      &32    \\  
  \noalign{\smallskip}
  \hline
  \end{tabular}
\tablebib{ (1) \citet{Gotz2006:08408-4503discovery}; (2) \citet{Mangano2011:atel3586}; (3) \citet{Walborn1973:08408};  (4) \citet[][]{Leyder2007}; 
(5) \citet{Lubinski2005}; (6) \citet{Romano2007}; (7) \citet{Vijapurkar1993};  (8) \citet{Masetti2006};  
(9)  \citet{Bird2007:igr3cat}; (10) \citet{Romano2013:Cospar12}; (11) \citet{Coleiro2013:IR_IDs}; (12) \citet{Fiocchi2012:16328-4726}; 
(13) \citet{Tomsick2004:atel224};  (14) \citet{Romano2012:sfxts_16418}; (15) \citet{Rahoui2008}; 
(16) \citet{Lutovinov2004:16465-4507}; (17) \citet{Walter2006};    (18) \citet{Negueruela2007}; (19) \citet{Nespoli2008};  
(20) \citet{Molkov2003:16479-4514};  (21) \citet{Ratti2010:chandrapos};   
(22) \citet{Smith1997:iauc6748}; (23) \citet{Smith1998:17391-3021}; (24) \citet{Smith2004:17391position}; (25) \citet{Negueruela2006};  
(26) \citet{Sunyaev2003}; (27) \citet{zand2005}; (28) \citet{Pellizza2006};   
(29) \citet{zand1998}; (30) \citet{zand2006:atel915}; (31) \citet{Negueruela2007:hmxbs}; (32) \citet{Torrejon2010:hmxbs};  
(33) \citet{Bamba1999:iauc7324}; (34) \citet{Bamba2001}; (35) \citet{Halpern2004:18410-0535a};  (36) \citet{Nespoli2008}; (37) \citet{Sguera2009};  
(38) \citet{Yamauchi1995:1845}; (39) \citet{Coe1996:18450};  
(40) \citet{Chernyakova2003}; (41) \citet{Giunta2009}. 
}
\end{sidewaystable*}
\setcounter{table}{1} 
 \begin{table*}
 \tabcolsep 3pt   	
 \caption{Sample of SFXTs:  Spin, orbital and superorbital periods, and proposed eccentricities. \label{sfxtcat:tab:datasprops} }   
 \begin{tabular}{llllllllll}   
 \hline
 \hline
\noalign{\smallskip}
Name                            &$P_{\rm spin}$                                        &$P_{\rm orb}$                                       &  $P_{\rm sup}$     & Eclipse & $e$ &  \multicolumn{4}{c}{Reference}          \\
                                     &(s)	                                                    &(d)                                                   &(d)	                 &              & &$P_{\rm spin}$ &$P_{\rm orb}$; $e$  &$P_{\rm sup}$ &Eclipse \\
  \noalign{\smallskip}
 \hline
 \noalign{\smallskip}
 IGR~J08408$-$4503   &  -                                   &$35$?                                                 &  -                               & N &  -                         & -    &1     &-    &-          \\  
IGR~J11215$-$5952   &$186.78\pm0.3$             &$164.6$                                              &  -                               & N & -                         &2     &3      &-    & -       \\ 
IGRJ~16328$-$4726 &  -                                      &$10.076\pm0.003$                            &  -                               & N & -                         & -    &4      &-    &-     \\
IGRJ~16418$-$4532 & $1209.12\pm0.42$          &$3.73886\pm0.00003$                      & $14.730\pm0.006$    & Y & -                         &5     &6      & 7    &  8   \\
IGR~J16465$-$4507  &$228\pm6$                      &$30.243\pm0.035$                            &  -                               & N & $<0.6$,$<0.8$    &9     &10;11     &-    &-       \\   
IGR~J16479$-$4514   &  -                                    &$3.3193\pm0.0005$                          &  $11.880\pm0.002$  & Y  & -                         & -    &12    &7    & 13       \\ 
XTE~J1739$-$302     &  -                                     &$51.47\pm0.02$\tablefootmark{a}    &  -                                & N & $<0.8$                & -   &14;14     &-    &-      \\ 
IGR~J17544$-$2619   &$71.49\pm0.02$             &$4.926\pm0.001$                              &  -                               & N  &$>0$                   & 15  &16;16      &-    &-      \\ 
SAX~J1818.6$-$1703 &  -                                    &$30\pm0.1$ 	                                  &  -                                & N  &$0.3$--$0.4$     & -   &17,18;18 &-   &-     \\ 
AX~J1841.0$-$0536   &$4.7394\pm0.0008$?\tablefootmark{b}  & -	                          &  -                                & N  &   -                      &19  &-        &-    &-    \\ 
AX~J1845.0$-$0433   &  -                                    & $5.7195\pm0.0007$	                  &  -                                & N  &$<0.37$             & -   &20;20  &-  &-    \\ 
IGR~J18483$-$0311   &$21.0526\pm0.0005$\tablefootmark{c}   &$18.545\pm0.003$ &   -                                & N  &$0.4$                 &21  &22;23      &-    & -    \\ 
  \noalign{\smallskip}
  \hline
  \end{tabular}
\tablefoot{
\tablefoottext{a}{See \citet[][$P_{\rm orb}=12.8658\pm0.0073$\,d]{Romano2009:sfxts_paperV}.}
\tablefoottext{b}{See \citet{Bozzo2011:18410}.}
\tablefoottext{c}{See \citet{Ducci2013:18483}.}
}
\tablebib{(1) \citet{Romano2009:sfxts_paper08408}; 
(2) \citet{Swank2007:atel999}; 
(3); \citet{Romano2009:11215_2008}; 
(4) \citet{Corbet2010:16328-4726}; 
(5) \citet{Drave2013:16418};  
(6) \citet{Levine2011}; 
(7) \citet[][]{Corbet2013:atel5126};  
(8) \citet{Corbet2006:atel779}; 
(9) \citet{Lutovinov2005};  
(10) \citet{LaParola2010:16465-4507_period}; 
(11) \citet{Clark2010:16465-4507_period2}; 
(12) \citet{Romano2009:sfxts_paperV};   
(13) \citet[][]{Bozzo2008:eclipse16479}; 
(14) \citet[][]{Drave2010:17391_3021_period}; 
(15) \citet{Drave2012:17544_2619_pulsation}; 
(16) \citet{Clark2009:17544-2619period}; 
(17) \citet{Bird2009:sax1818.6_period};  
(18) \citet{Zurita2009:sax1818.6_period}; 
(19) \citet{Bamba2001}; 
(20) \citet{Goossens2013:18450_period};  
(21) \citet{Levine2011};  (22) \citet{Levine2006:igr18483}; (23) \citet{Romano2010:sfxts_18483}.
}
\end{table*}

	\section{The SFXT Sample} \label{sfxtcat:sample}

Defining a complete sample of SFXTs has been a challenge, due to the initially 
 loose constraints often applied to the defining characteristics of their X--ray 
emission and the intrinsic difficulty in performing optical spectroscopy on the 
often heavily absorbed companion stars. 
In this work, we distinguish between confirmed and SFXT candidates 
based on the availability of an optical classification of the companion:  
a {\it confirmed} SFXT is a transient that has shown a repeated,  
high-dynamical range, flaring activity that is firmly associated with an OB supergiant, 
while an SFXT {\it candidate} has shown similar X--ray behaviour 
but has no confirmed association with an OB supergiant companion. 

The sources in our sample of SFXTs were selected from the literature 
based on evidence of bright flares (peak $L\ga10^{36}$ erg s$^{-1}$), as  recorded by ASCA, \rxte, 
\inte,  and \sw. 
The full list of all confirmed SFXTs that triggered the BAT 
is reported in Table~\ref{sfxtcat:tab:datapositions} 
along with the currently most accurate X--ray positions, their errors  (Col.\ 2, 3, and 4), 
the optical counterpart, its spectral type, distance (Col.\ 5, 6, and 7), 
the reference to discovery, positions, and counterpart properties (Col.\ 8, 9, 10, and 11).    
Table~\ref{sfxtcat:tab:datasprops} reports the spin and orbital periods, eccentricities, super-orbital periods
and the presence of eclipses in the X--ray light curve. 

In the following,  we summarize the basic information on each source in the SFXT 
class we considered for this work  
with a special emphasis on those that triggered the BAT (in any of the ways described above).

\subsection{IGR~J08408$-$4503 \label{sfxtcat:obj:08408}} 
The transient \object{IGR~J08408$-$4503} was discovered on 2006 May 15 during a 900\,s  bright flare that  
reached a peak flux of 250 mCrab \citep[20--40\,keV, ][]{Gotz2006:08408-4503discovery}. 
\citet{Mereghetti:08408-4503} later demonstrated its recurrent transient nature by discovering  
an earlier active state in 2003. 
It is associated with the O8.5Ib(f) supergiant star, \object{HD~74194}, 
in the Vela region \citep{Masetti2006:08408-4503} at a distance of $\sim 3$\,kpc 
\citep{Leyder2007}. 
No information is available on the orbital or spin periods.  

IGR~J08408$-$4503 triggered the BAT several times with its 
most remarkable  outburst on 2008 July 5 \citep{Romano2009:sfxts_paper08408}, during which 
the XRT light curve showed a multiple-peaked structure with a first bright 
flare that reached $\sim 10^{-9}$ erg\,cm$^{-2}$\,s$^{-1}$ (2--10\,keV), which were 
followed by two equally bright flares within 75\,ks. 
The spectral characteristics of these flares differ dramatically, 
with most of the difference being due to absorbing column variations. 
A gradual decrease of the $N_{\rm H}$ was also observed and interpreted as 
due to an ionization effect produced by the first flare, which resulted in a significant 
decrease in the measured column density towards the source. 

Recent papers report the properties of the quiescent state of IGR~J08408$-$4503. 
\citet[][]{Bozzo2010:quiesc1739n08408}, in particular, show that the flaring behaviour extends 
down to $3\times10^{-13}$ erg\,cm$^{-2}$\,s$^{-1}$ (0.5--10\,keV, or a 
luminosity of $3.3\times10^{32}$ erg\,s$^{-1}$), thus confirming a dynamical range in 
excess of $10^{4}$.

\subsection{IGR~J11215$-$5952  \label{sfxtcat:obj:11215}} 
The object \object{IGR~J11215$-$5952} was discovered on 2005 April 22 when it reached $\sim 75$\,mCrab (20--60 keV). 
It was  associated with \object{HD~306414} \citep{Negueruela2005b}, 
a B1Ia  supergiant \citep{Vijapurkar1993} located at a distance of 6.2~kpc \citep{Masetti2006}.  
The source is a pulsar with spin period $P_{\rm spin}=186.78\pm0.3$\,s \citep{Swank2007:atel999}. 
The orbital period, which was initially supposed to be $\sim 330$\,d 
\citep[][]{SidoliPM2006}, was later pinpointed, by means of \sw\ 
\citep[][]{Romano2007apastron,Sidoli2007,Sidoli2008:atel1444,Romano2009:11215_2008}, 
to $P_{\rm orb}=164.6$\,d,  the longest measured for an SFXT. 

This source has been the focal point of several \sw/XRT observing campaigns, starting from the 
observations of the 2007 February 9 outburst \citep[][]{Romano2007}, through which
we discovered that the X--ray emission, hence the accretion phase, lasts much longer 
than previously observed using less sensitive instruments.
The timescale is not on the order of minutes or hours, but several days. 
In particular, in \citet{Sidoli2007} an explanation was proposed for the SFXTs outburst  
based on the periodic nature of the outbursts of this source and on the narrow shape 
of its X--ray light curve as observed with XRT.  The periodicity in the outbursts suggests
it is being driven by the orbital period, with outbursts triggered at, or near, the periastron passage.
The shape of the observed light curve, on the other hand, can only be explained with 
accretion from non spherically-symmetric winds. Therefore, the suggestion was made of the presence of a second 
component of the clumpy wind, such as an equatorially-enhanced wind component 
(or any other preferential plane for the outflowing wind), 
which is denser and slower than the symmetric polar wind from the blue supergiant, 
and inclined with respect to the orbital plane of the system.
This object  has also recently been added to the new class of Galactic transient 
MeV/TeV emitters, as a counterpart candidate of \object{EGR~J1122$-$5946} \citep{Sguera2008:11215}.

We note that IGR~J11215$-$5952 never triggered the BAT, so it is only included in this Section 
for the sake of completeness and will not be discussed further.

\subsection{IGR~J16328$-$4726   \label{sfxtcat:obj:16328}} 

The source \object{IGR~J16328$-$4726} \citep{Bird2007:igr3cat} 
has a history of hard X-ray activity characterized by short flares 
lasting up to a few hours \citep{Fiocchi2010:16328-4726}. 
It  triggered the \sw/BAT on 2009 June 10 \citep{Romano2013:Cospar12}, 
when the source reached an unabsorbed 2--10\,keV flux of $\sim 4\times10^{-10}$ 
erg cm$^{-2}$ s$^{-1}$. The \sw\ arcsecond position allowed 
\citet{Grupe2009:16328-4726} to propose the IR star 
\object{2MASS~J16323791$-$4723409} as the optical counterpart which is 
classified as an O8Iafpe supergiant star by \citet{Coleiro2013:IR_IDs}. 
The orbital period is $P_{\rm orb}=10.076\pm0.003$\,d \citep{Corbet2010:16328-4726}. 

Recently, \citet[][]{Bozzo2012:HMXBs} reported  luminosity variations by a factor of 10
during a 22\,ks \xmm\ observation performed on 2011 February 20, when 
the source was in a much fainter  flux state 
(unabsorbed $F_{\rm 2-10\,keV}=1.7\times10^{-11}$  
 erg cm$^{-2}$ s$^{-1}$) than the one observed during the BAT outburst.

 \subsection{IGRJ~16418$-$4532   \label{sfxtcat:obj:16418}} 
The transient \object{IGR~J16418$-$4532} was discovered on 2003 February 1--5 \citep{Tomsick2004:atel224} 
during an \inte\ scan of the Norma Region when it reached a flux of 
$3\times10^{-11}$ erg cm$^{-2}$ s$^{-1}$ (20--40\,keV). 
Subsequent flares were reported by \citet{Sguera2006} on 2004 February 26, 
during which fast ($\sim 1$\,hr) X-ray outbursts were observed peaking at 
$\sim 80$\,mCrab (20--30\,keV).  
\citet{Ducci2010} detected 23 more outbursts and calculated an activity duty cycle of 
$\sim 1$\,\% (one of the highest among the 14 SFXTs and SFXT candidates 
they examined) and fluxes ranging between $1.3\times10^{-10}$ erg cm$^{-2}$ s$^{-1}$ 
and $4.8\times10^{-10}$ erg cm$^{-2}$ s$^{-1}$ (18--100\,keV). 
Based on the \xmm\ position,  \citet{Chaty2008} proposed  
\object{2MASS~J16415078$-$4532253} as the best near infrared (NIR) counterpart candidate, 
which is classified as BN0.5Ia by \citet{Coleiro2013:IR_IDs}. 

\citet{Walter2006} discovered a pulsation at $1246\pm100$\,s.  
 \citet{Corbet2006:atel779} discovered an orbital period of $\sim 3.75$\,d,  
as  based on \rxte/ASM data, and an eclipse. 
Recently, a super-orbital modulation has been detected in the \sw/BAT light curve 
\citep{CorbetKrimm2013:superorbital}  
at $P_{\rm sup}=14.730\pm0.006$\,d. 

This source was intensively observed with \sw/XRT along the orbital period between 
2011 July 13--30 \citep{Romano2012:sfxts_16418}. 
As observed in IGR~J18483$-$0311, the light curve shows an orbital modulation and flaring episodes. 
By assuming a circular orbit, \citet{Romano2012:sfxts_16418} could explain their X-ray emission in terms of 
the accretion from a spherically symmetric clumpy wind,  which is composed of clumps with masses 
ranging from $\sim5\times10^{16}$~g to $10^{21}$~g.

 \subsection{IGR~J16465$-$4507  \label{sfxtcat:obj:16465}} 
The source \object{IGR~J16465$-$4507} was discovered by \inte\ on 2004 September 6--7 \citep[][]{Lutovinov2004:16465-4507}, 
when it averaged  $8.8\pm0.9$\,mCrab (18--60\,keV) and subsequently showed a strong flare at $\sim 28$\,mCrab
on September 7. The optical counterpart \object{2MASS J16463526$-$4507045} 
\citep[][]{ZuritaHeras2004:16465-4507} 
was classified as a B0.5I star by \citet[][USNO-B1.0 0448$-$00520455]{Negueruela2005:16465-4507} 
and later refined to B0.5Ib \citep{Negueruela2007}  at a distance of about 8\,kpc 
\citep[but also see][]{Nespoli2008,Rahoui2008}. 

It is a pulsar with $P_{\rm spin}=228\pm6$\,s \citep{Lutovinov2005}
and orbital period $P_{\rm orb}=30.243\pm0.035$\,d \citep{LaParola2010:16465-4507_period}, 
and  also never triggered the \sw/BAT.

 \subsection{IGR~J16479$-$4514  \label{sfxtcat:obj:16479}} 
The object \object{IGR~J16479$-$4514} was discovered on 2003 August 8--9 
\citep{Molkov2003:16479-4514} during an outburst that reached $\sim12$\,mCrab (18--25\,keV) 
and $\sim8$\,mCrab (25--50\,keV).  
Since then the source has shown frequent flaring activity, as recorded by both 
\inte\ \citep[][]{Sguera2005,Sguera2006,Walter2007} and \sw{} 
\citep[][]{Kennea2005:16479-4514,Markwardt2006:16479-4514,Romano2008:sfxts_paperII,
Bozzo2009:16479-4514outburst,Romano2009:sfxts_paperV}. 
It is associated with an O8.5I star at a distance of 4.9\,kpc \citep{Rahoui2008}. 

The orbital period $P_{\rm orb}=3.3194\pm0.0010$\,d   was 
discovered by \citet{Jain2009:16479_period} 
in the first 4 years of \sw/BAT data and \rxte/All Sky Monitor (ASM) data, 
and it is the shortest measured for an SFXT.

\citet{Bozzo2008:eclipse16479}  reported an X--ray eclipse by the supergiant companion 
observed during a long \xmm\ observation obtained after the 2009 March 19 outburst.  
 Recently, a super-orbital modulation has been detected in the \sw/BAT light curve 
\citep{CorbetKrimm2013:superorbital}   
at $P_{\rm sup}=11.880\pm0.002$\,d.

 \subsection{XTE~J1739$-$302: the SFXT class prototype     \label{sfxtcat:obj:17391}} 
The transient \object{XTE~J1739$-$302} was discovered in August 1997  by \rxte\ \citep{Smith1997:iauc6748,Smith1998:17391-3021}, 
when it reached a peak flux of 3.6$\times$10$^{-9}$~erg cm$^{-2}$~s$^{-1}$ (2--25 keV), 
and has a long history of flaring activity recorded by \inte\ \citep{Sguera2006,Walter2007,Blay2008}  
and by \sw\ \citep{Sidoli2009:sfxts_paperIII,Sidoli2009:sfxts_paperIV,Romano2011:sfxts_paperVI,Farinelli2012:sfxts_paperVIII}. 
It is now considered a prototype of the SFXT class. 
The optical counterpart is an O8Iab(f) star \citep{Negueruela2006} at a distance of 
2.7\,kpc \citep{Rahoui2008}.   

 \citet{Drave2010:17391_3021_period} reported the discovery of a $P_{\rm orb}=51.47\pm0.02$\,d 
orbital period based on $\sim 12.4$\,Ms of \inte\ data. We note however that this period was not 
independently confirmed by an \rxte\ investigation \citep{Smith2012:FXRT_RXTE} and a hint 
of a periodicity at $P_{\rm orb}=12.8658\pm0.0073$\,d  
(1/4 of the  value above)
was reported by \citet[][]{Romano2009:sfxts_paperV}. 

 The properties of the quiescent state of  XTE~J1739$-$302 are described in \citet[][]{Bozzo2010:quiesc1739n08408}. 
As observed in  IGR~J08408$-$4503, the flaring behaviour extends down to 
$4.7\times10^{-13}$ erg\,cm$^{-2}$\,s$^{-1}$ (0.5--10\,keV, or a luminosity of $4.1\times10^{32}$ erg\,s$^{-1}$).

 \subsection{IGR~J17544$-$2619: the SFXT class prototype  \label{sfxtcat:obj:17544 }} 
The first 2-hr flare from \object{IGR~J17544$-$2619} was observed by \inte\ on 2003 September 17 \citep{Sunyaev2003}, 
when the source reached a flux of 160~mCrab (18--25~keV). 
Several more flares, lasting up to 10 hours, were detected by \inte\ in  the following years 
\citep{Grebenev2003:17544-2619,Grebenev2004:17544-2619,Sguera2006,Walter2007,Kuulkers2007} 
with fluxes up to 400~mCrab (20--40 keV), and some were found in archival \sax\ observations 
\citep{zand2004:17544bepposax}. 
Subsequent flares were observed by \sw\ 
\citep{Krimm2007:ATel1265,Sidoli2009:sfxts_paperIII,Sidoli2009:sfxts_paperIV,Romano2011:sfxts_paperVI,Romano2011:sfxts_paperVII,Farinelli2012:sfxts_paperVIII} 
and \suzaku\ \citep[][which caught a $\ga$ day long outburst]{Rampy2009:suzaku17544}, 
so that IGR~J17544$-$2619 is now considered a prototype of the SFXT class.  
The optical counterpart is an O9Ib star \citep{Pellizza2006} at 3.6\,kpc \citep{Rahoui2008}. 

 \citet{Clark2009:17544-2619period} reported the discovery of a $4.926\pm0.001$\,d 
orbital period based on the $\sim 4.5$ years of \inte\ data, while 
\citet[][]{Drave2012:17544_2619_pulsation} reported a pulsation at $71.49\pm0.02$\,s 
from the region around the source that they attribute to a spin period. 

 The first detailed observations of an SFXT in quiescence were performed on
this source. \citet{zand2005} reported  a {\it Chandra} observation where 
the source is characterized by a very soft ($\Gamma=5.9\pm1.2$)  quiescent 
($L\sim5\times10^{32}$ erg\,s$^{-1}$) spectrum. 
The initial state of quiescence is then followed by a bright outburst, implying a 
dynamical range of at least 4 orders of magnitude in observed flux. 
These observations prompted \citet[][]{Bozzo2008} to interpret the 
very large luminosity ranges observed on timescales as short as hours as 
transitions across the magnetic and/or centrifugal barriers, thus
envisioning a scenario in which SFXTs with large dynamic range and 
large $P_{\rm spin} \ga 1000$\,s are characterized by 
magnetar-like fields ($B\ga 10^{14}$\,G).

 \subsection{SAX~J1818.6$-$1703 \label{sfxtcat:obj:1818}} 
The source \object{SAX~J1818.6$-$1703} was discovered on 1998 March 10--12 
with the Wide Field Cameras on board \sax\ \citep{zand1998} 
as a hard transient that reached 100\,mCrab in the 2--9\,keV band and
400\,mCrab in 9--25\,keV band. 
Several more bright flares lasting 1--3 hours were observed \citep[][]{Grebenev2005,Sguera2005} 
with IBIS/ISGRI on board \inte\ reaching $\sim$200 mCrab 
at the flare peak (18--45 keV), and with \rxte/ASM \citep{Sguera2005}. 
\citet{Bozzo2008:atel1493} estimated a dynamic range of 4 orders of magnitude. 
Further activity has been caught both by \inte\ \citep{Grebenev2008:atel1482} 
and  \sw\ \citep{Sidoli2009:sfxts_sax1818,Romano2009:atel2191,Romano2009:atel2279}.
\citet{Negueruela2006:aTel831} proposed an association with 
\object{2MASS~J18183790$-$1702479} (USNO-B1.0 0729$-$0750578), 
a supergiant star with an earlier than B3 spectral type, 
which was  later confirmed by a {\it Chandra} observation \citep{inzand2006:aTel915}. 
The spectral type was refined to O9-B1I \citep{Negueruela2007:hmxbs} 
and then to B0.5Iab at a distance of $2.1\pm0.1$\,kpc \citep{Torrejon2010:hmxbs}. 

\citet{Bird2009:sax1818.6_period} and \citet{Zurita2009:sax1818.6_period}  
discovered an orbital periodicity of $30\pm0.1$\,d 
from the analysis of available \sw/BAT and \inte\ data, suggesting an eccentric 
orbit ($e$$\sim$0.3--0.4) and a typical outburst duration of 4--6~days.

\subsection{AX~J1841.0$-$0536 \label{sfxtcat:obj:18410}} 
The object \object{AX~J1841.0$-$0536} was discovered during {\it ASCA} observations of the Scutum arm region 
that were performed on 1994 April 12 and 1999 October 3--4 as a flaring source,  
which exhibited 
flux increases by a factor of 10 (up to $\sim 10^{-10}$ erg cm$^{-2}$ s$^{-1}$)
with rise times on the order of 1\,hr  \citep{Bamba1999:iauc7324,Bamba2001}, 
a strong absorption $N_{\rm H} =3\times10^{22}$ cm$^{-2}$,  
and coherent pulsations with a period of $4.7394\pm0.0008$\,s 
\citep[but also see][]{Bozzo2011:18410}.  
\citet[][]{Rodriguez2004:18410-0535} later discovered IGR~J18410$-$0535, which 
was observed to flare by \inte\ on 2004 October 8 and 
reached $\sim 70$\,mCrab (20--60\,keV)  
and 20\,mCrab (60--200\,keV) and was subsequently identified with AX~J1841.0$-$0536
\citep{Halpern2004:18410-0535b}. 
The infrared (IR) counterpart is \object{2MASS 18410043$-$0535465}, a reddened star with a 
weak double-peaked  H$\alpha$ emission line, which was initially classified as a Be star, but 
later reclassified as B1Ib type star \citep{Nespoli2008}.  
This corroborated the evidence that AX~J1841.0$-$0536 is a member of the SFXT class,
as proposed by \citet{Negueruela2006:ESASP604}, at a distance of $3.2_{-1.5}^{+2.0}$\,kpc
\citep{Nespoli2008}. 

 Several flares have been seen by \inte\ \citep{Sguera2006,Sguera2009}, 
MAXI \citep[][]{negoro2010:atel3018},  
\sw\ \citep[e.g.][]{Romano2011:sfxts_paperVII}, and \xmm\ \citep[][]{Bozzo2011:18410}. 
In particular, \citet{Bozzo2011:18410} report evidence that the flare observed 
was produced by the accretion of a massive clump onto the compact object hosted in 
this SFXT. 

 Recently, AX~J1841.0$-$0536 has been proposed as the prototype of a new class of Galactic transient 
MeV/TeV emitters due to it being a possible counterpart of \object{3EG~J1837$-$0423} \citep{Sguera2009}.

\subsection{AX~J1845.0$-$0433 \label{sfxtcat:obj:18450}} 
 The source \object{AX~J1845.0$-$0433} 
was discovered in ASCA data \citep[][]{Yamauchi1995:1845} as a source 
variable on timescales of tens of minutes \citep{Sguera2007:18450,Zurita2009:1845}
and was classified as a SFXT with an O9.5I companion 
at a distance of 3.6\,kpc \citep[][]{Coe1996:18450}. 
 It also triggered the \sw/BAT 
\citep[][]{Romano2009:atel2102,Romano2012:atel4095,Romano2013:Cospar12}. 
 Recently, a modulation in the light curve has been observed in \inte\ data 
at  $5.7195\pm0.0007$\,d \citep[][]{Goossens2013:18450_period}
and attributed to an orbital period.

 \subsection{IGR~J18483$-$0311 \label{sfxtcat:obj:18483}} 
The transient \object{IGR~J18483$-$0311} was first detected during observations of the Galactic Centre 
on 2003 April 23--29, when it reached a flux of 10\,mCrab in the 15--40\,keV energy band and 
5\,mCrab in the 40--100\,keV band \citep{Chernyakova2003}.  
Several flares were subsequently observed by \inte\ \citep{Sguera2007}. 
Some of them exceeded one day in length and reached  a peak flux of 120\,mCrab. 
The source was first associated with a B0.5Ia star, located at a distance of 3--4\,kpc 
\citep{Rahoui2008:18483}, a classification later refined by \citet{Torrejon2010:hmxbs} to 
B0-B1Iab at $2.83\pm0.05$\,kpc. 
It is probably the most active SFXT as observed by 
MAXI\footnote{See the MAXI notices at 
\href{http://maxi.riken.jp/pipermail/x-ray-star/}{http://maxi.riken.jp/pipermail/x-ray-star/. }
}.

An orbital period of $18.55\pm{0.03}$\,d  was  
discovered by \citet{Levine2006:igr18483} 
in the \rxte/ASM data.  
\citet{Sguera2007}  discovered pulsations at $P_{\rm spin}=21.0526\pm{0.0005}$\,s 
with the X--ray monitor JEM-X  (but see \citealt{Ducci2013:18483}).

So far, IGR~J18483$-$0311 only triggered the \sw/BAT once, and in that case, no NFI data were 
gathered. However, it was intensively observed with \sw/XRT between 2009 June 11 and July 9 
\citep{Romano2010:sfxts_18483} along the orbital period. 
The XRT light curve shows an orbital modulation and the flaring as characteristic of SFXTs. 
By assuming an eccentricity of $e = 0.4$, \citet{Romano2010:sfxts_18483}  
could explain their X-ray emission in terms of 
the accretion from a spherically symmetric clumpy wind, composed of clumps with different masses
which range from $10^{18}$ to $5\times10^{21}$\,g.

\setcounter{table}{2} 
 \begin{table} 	
\tabcolsep 3pt         
\caption{On-board triggers and detections throughout the \sw\ mission (2005-02-12 to 2013-05-31). 
The flag denotes the subsample:  T = BAT trigger, 
     D = daily-averaged BATTM lightcurves, 
     o =  orbital-averaged BATTM lightcurves, and 
     d =  on-board detections, as described in Sect.~\ref{sfxtcat:batdetections}. 
The numbers in parenthesis in the T, D, and o columns and in the Totals line 
report the total number of days the source was detected 
(when more than one detection was achieved in a given day, the brightest detection was selected, see Sect.~\ref{sfxtcat:batdetections}).  
\label{sfxtcat:tab:alltriggers}} 	
 \centering
 \begin{tabular}{lllrr} 
 \hline 
 \hline 
 \noalign{\smallskip} 
Name                         & BAT 	                        & BATTM 	         &BATTM                  &  BAT	    \\
                                  & on-board 	                &$>5 \sigma$ 	 &$>5 \sigma$        &  on-board	    \\
                                  &  triggers 	                & daily	                 & orbital &  detections \\   
\hspace{2truecm}Flag                            & T	                        & D	                         & o	                          & d \\    
 \noalign{\smallskip} 
 \hline 
 \noalign{\smallskip} 
IGR~J08408$-$4503  &	7 (6)\tablefootmark{a}	 &	4	&	7 (5)   & 50 (8)	\\
IGR~J16328$-$4726 &	2	                         &	0	&	4 (3)   & 4 (2)	\\
IGR~J16418$-$4532 &	3	                         &	5	&	17 (16) & 19 (10)	\\
 IGR~J16465$-$4507  &	0	                         &	1	&	1 (1)   & 1 (1) 	        \\
IGR~J16479$-$4514  &	8 (7)\tablefootmark{a}	 &	39	&	75 (61) & 147 (50)	\\
XTE~J1739$-$302 	  &	8 (7)\tablefootmark{a}	 &	5	&	39 (29) & 124 (37) 	\\
IGR~J17544$-$2619  &	5	                         &	12	&	32 (23) & 90 (22)	\\
SAX~J1818.6$-$1703&	5	                         &	8	&	23 (17) & 54 (14)	\\
AX~J1841.0$-$0536 &	4	                         &	8	&	24 (16) & 48 (17)	\\
AX~J1845.0$-$0433 &	3	                         &	3	&	11 (8) & 17 (9)	\\
IGR~J18483$-$0311 &	1	                         &	41	&	34 (24) & 124 (35)	\\
   \noalign{\smallskip}
  \hline
   \noalign{\smallskip}
 Totals                        &    46 (43)                     & 126     &      267 (203) & 678 (205) \\ 
 \noalign{\smallskip}
  \hline
  \end{tabular}
\tablefoot{
\tablefoottext{a}{The source triggered the BAT twice within a few hours.}
}
   \end{table}

\setcounter{table}{3} 
 \begin{table*}
 \tabcolsep 4pt         
 \caption{First 15 entries of the 100-month {\it Swift} catalogue of SFXTs.\label{sfxtcat:tab:catalogue} }
 \centering
\begin{tabular}{llccrcccrrrc} 
 \hline 
 \noalign{\smallskip} 
Num &    Name                       &    Flag   &      Year         & Doy     &      MJD                          &     UT Date              &     UTtime        & Duration             & S/N        & Flux     &Trigger  \\ 
              &                                    &             &                      &            &                                       &    &       & (s)                      &           & (mCrab)     &\#  \\ 
\noalign{\smallskip} 
 \hline 
 \noalign{\smallskip} 
1   & IGRJ08408-4503 & D & 2008 & 265 & 54730.00000 & 2008-09-21 & 00:00:00 & 86400.0 & 6.85 & 17.0 & 999999 \\
2   & IGRJ08408-4503 & D & 2009 & 137 & 54968.00000 & 2009-05-17 & 00:00:00 & 86400.0 & 5.28 & 18.0 & 999999 \\
3   & IGRJ08408-4503 & D & 2009 & 240 & 55071.00000 & 2009-08-28 & 00:00:00 & 86400.0 & 5.50 & 31.0 & 999999 \\
4   & IGRJ08408-4503 & D & 2011 & 237 & 55798.00000 & 2011-08-25 & 00:00:00 & 86400.0 & 15.55 & 69.0 & 999999 \\
5   & IGRJ08408-4503 & T & 2006 & 277 & 54012.61328 & 2006-10-04 & 14:45:42 & 1600.0   & 8.08 & 99999.0 & 232309 \\
6   & IGRJ08408-4503 & T & 2008 & 187 & 54652.88672 & 2008-07-05 & 21:14:13 & 64.0       & 7.38 & 152.0 & 316063 \\
7   & IGRJ08408-4503 & T & 2008 & 265 & 54730.32812 & 2008-09-21 & 07:55:08 & 64.0       & 10.00 & 391.0 & 325461 \\
8   & IGRJ08408-4503 & T & 2009 & 240 & 55071.95312 & 2009-08-28 & 22:51:46 & 320.0     & 6.62 & 1930.0 & 361128 \\
9   & IGRJ08408-4503 & T & 2009 & 240 & 55071.96484 & 2009-08-28 & 23:09:22 & 64.0       & 10.26 & 99999.0 & 361129 \\
10 & IGRJ08408-4503 & T & 2010 &   87 & 55283.66406 & 2010-03-28 & 15:53:38 & 64.0      & 8.68 & 170.0 & 417420 \\
11 & IGRJ08408-4503 & T & 2011 & 237 & 55798.03516 & 2011-08-25 & 00:53:04 & 64.0       & 7.28 & 203.0 & 501368 \\
12 & IGRJ08408-4503 & d & 2006 & 118 & 53853.22266 & 2006-04-28 & 05:23:18 & 112.0     & 5.60 & 101.0 & 999999 \\
13 & IGRJ08408-4503 & d & 2006 & 277 & 54012.61328 & 2006-10-04 & 14:45:42 & 312.0     & 6.28 & 102.0 & 999999 \\
14 & IGRJ08408-4503 & d & 2006 & 277 & 54012.61328 & 2006-10-04 & 14:45:42 & 1600.0   & 8.09 & 82.0 & 999999 \\
15 & IGRJ08408-4503 & d & 2006 & 277 & 54012.62109 & 2006-10-04 & 14:56:14 & 320.0    & 5.90 & 94.0 & 999999  \\
  \noalign{\smallskip}
  \hline
  \end{tabular}
\tablefoot{The full catalogue is available at CDS. Unvailable fluxes are set to 99999.0\,mCrab, 
and flares that are not BAT triggers are arbitrarily assigned 999999 as trigger number. \\
 }
\end{table*}

 	 \section{Analysis and results} \label{sfxtcat:dataredu} %

 	 \subsection{BAT data subsamples} \label{sfxtcat:batdetections}

For each source in the  BAT Transient Monitor (BATTM), the data products are two light curves: 
daily average and orbit-level (averaging $\sim 800$\,s). 
\citet{Krimm2013:BATTM} consider a source detected if it meets either 
of the following criteria:  the mean rate has a value $M\ga3$\,mCrab,  and  
the peak rate for days when the source was found at $\ga 7\sigma$ to be 
$P_{7}\ga 30$\,mCrab. 
New sources are generally announced to the astronomical community  through GCN  if they are observed 
at $\ga 6\sigma$ for two or more days in the 1-day mosaics or at $\ga 6\sigma$ in a 
multi-day mosaic. The announcement is automatic whenever a source reaches $\ga 8\sigma$. 
SFXTs are `known' sources; therefore,  no automatic announcements are  generally made but 
an Astronomer's Telegram\footnote{
\href{http://www.astronomerstelegram.org}{http://www.astronomerstelegram.org .}
} is issued upon examination of interesting events. 

For this paper, we have collected all detections in excess of $5\sigma$ from both 
daily- and orbital-averaged BATTM light curves in the time range of 
2005 February 12 to 2013 May 31 (MJD~53413--56443). 
Hereafter, these two subsamples are denoted by (D) and (o), respectively. 

\begin{figure} 
\includegraphics[width=\hsize]{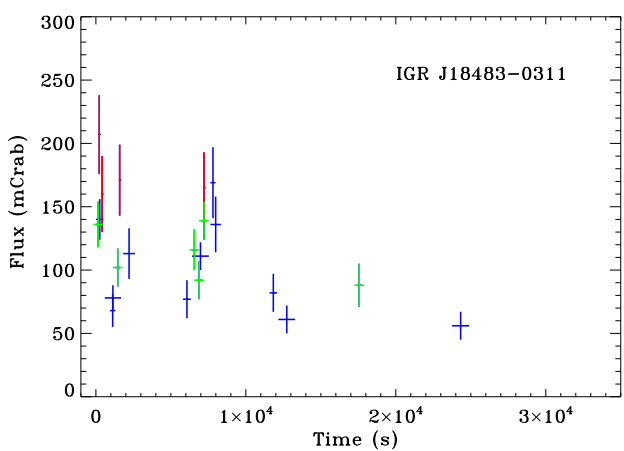}
\caption{Example of on-board data: 15--50\,keV light curve of IGR~J18483$-$0311. 
The red points are drawn from 64\,s images, the green ones from 320\,s images,
and the blue ones from other (generally longer) timescale images. 
} 
\label{sfxtcat:fig:example_d} 
\end{figure}

As described in \citet[][]{Fenimore2003:BATtrigger_algorithm}, the BAT on-board  
trigger algorithm works on several different timescales. 
The triggering code has three types of triggers: 
two are based on increases of count rates (short rate triggers on timescales of 4--64\,ms; 
long rate triggers on timescales of 64\,ms--24\,s), 
and one is based on images (image triggers, on timescales of 64\,s to many minutes). 
In the latter case, each image is searched for significant sources, but \sw\ 
does not slew to known sources unless the image flux exceeds a threshold 
set in the on-board source catalog.  The on-board source thresholds are set high in most 
cases but are manually set to low values for SFXTs, so that \sw\ will slew to them when they become active.
An example of the on-board data is shown in Fig.~\ref{sfxtcat:fig:example_d},
which shows the 15--50\,keV light curve of IGR~J18483$-$0311 during a very active day, 
2007 September 21.  
The red points are derived from 64\,s images, the green ones from 320\,s images, and 
the blue ones from other (generally longer) timescale images that are 
generated and analysed on-board by BAT when it is not responding 
to detected rate increases. 
The superposition of the times is due to the image detection algorithm. 

For this work, we considered all on-board detections ($\ga 5\sigma$) as 
recorded  in the same time span and selected those within 4 arcmin of each source 
in our sample. Hereafter, this subsample is denoted by (d).

Given the cut in $\sigma$ applied to the available BATTM and on-board detections, 
our catalogue is a flux limited sample of flares. 
Assuming a Crab-like spectrum (power-law of photon index 2.15),
$5\sigma$ detections for one day and an average orbit typically correspond to fluxes of 
$5.98\times10^{-10}$ and  $1.46\times10^{-9}$ erg cm$^{-2}$ s$^{-1}$, respectively, 
in the 15--150\,keV band (or $3.38\times10^{-10}$ and  $8.24\times10^{-10}$ 
erg cm$^{-2}$ s$^{-1}$ in the 15--50\,keV band).  
We note that  the Crab produces a count rate 
of 0.22 counts cm$^{-2}$ s$^{-1}$ in the BAT (15--50\,keV band, on-axis) 
or a flux of $1.3\times10^{-8}$ erg cm$^{-2}$ s$^{-1}$.

Table~\ref{sfxtcat:tab:alltriggers} summarizes our results and reports for each source 
the total number of detections in the daily- and orbital-averaged 
light curves (Cols.~3 and 4, marked as ``D'' and ``o'') 
and the on-board detection (Col.~5, ``d'').  
As typical of SFXTs, for a given source several detections occur on the same date;  
therefore, we also provide the total number of days the source was 
detected in the orbital-averaged light curves and on-board 
in parenthesis in Col.~4 and 5 of Table~\ref{sfxtcat:tab:alltriggers}.  
For completeness, we also report the  total number of BAT triggers, 
that is the instances when the significance was high enough to trigger the 
BAT and initiate the typical \sw\ GRB follow-up response (Col.~2,  ``T'').

In Tables~\ref{sfxtcat:tab:08408dets}--\ref{sfxtcat:tab:18483dets},  
we report the results on individual sources, as an abridged list of detections--we 
selected the brightest flare for each day the source was detected--as marked by an (o) and (d), 
for orbital-averaged  and  on-board detection, respectively. 
The detections from the daily-averaged BATTM light curves are marked by a (D). 
In Tables~\ref{sfxtcat:tab:08408dets}--\ref{sfxtcat:tab:18483dets},    
we also report the subsample of BAT triggers, marked by a (T), along with the 
trigger number (Col.~6).

The full catalogue file contains the following fields: 
unique line identifier (Col.~1);  
source name (Col.~2); 
a flag (Col.~3) discriminating the detection method (T=BAT trigger, 
     D=from daily-averaged BATTM light curves, 
     o=from orbital-averaged BATTM light curves, 
     d=from on-board detections); 
 year, day of year (DOY) and MJD (Cols.~4, 5, and 6); 
 UT date and time (Cols.~7, 8); 
 flare duration (Col.~9); 
significance of the detection in units of $\sigma$ (Col.~10); 
mean flux in mCrab  (Col.~11); 
 and trigger number (Col.~12 for BAT triggers).  
The first 15 lines are reported in Table~\ref{sfxtcat:tab:catalogue}.

\begin{figure} 
\includegraphics[width=\hsize]{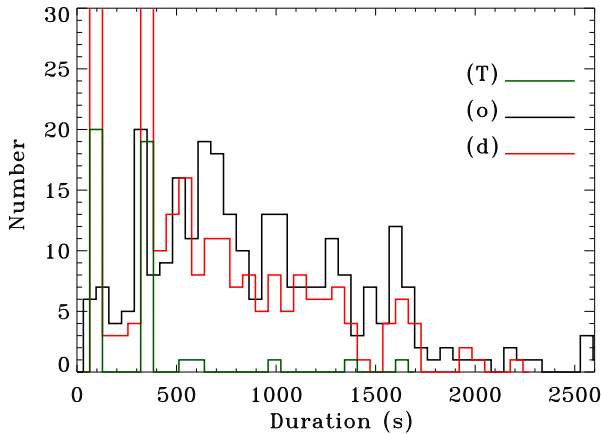}
\caption{Distributions of the flare durations for BAT triggers (T),
orbital-averaged BATTM light curves (o), and 
on-board detections (d) in units of seconds, drawn from the full catalogue. 
The peaks at 64 and 320\,s reach $\sim 270$ and 250, respectively. 
} 
\label{sfxtcat:fig:durations} 
\end{figure} 

 	 \subsection{Flare ensemble properties} \label{sfxtcat:statistics}

Our sample comprises a total of 1117 flares from 11 sources. 
They are divided into 
46 BAT triggers (T, in 43 outbursts),
126 daily-averaged BATTM light curves (D),
267 orbital-averaged BATTM light curves (o), and 
678 on-board detections (d). 

Figure~\ref{sfxtcat:fig:durations} shows the distributions of the observed durations. 
The means are 285\,s (T), 897\,s  (o), 351\,s  (d);  and 
their medians are 320\,s (T) and 792\,s   (o). 
We note that most on-board detections of SFXT flares result from the (nominally) 
64\,s, 320\,s, and full-pointing images that BAT generates. 
The image duration is used as a proxy for the flare duration, leading 
to the two peaks in Figure~\ref{sfxtcat:fig:durations}.

\begin{figure} 
\includegraphics[width=\hsize]{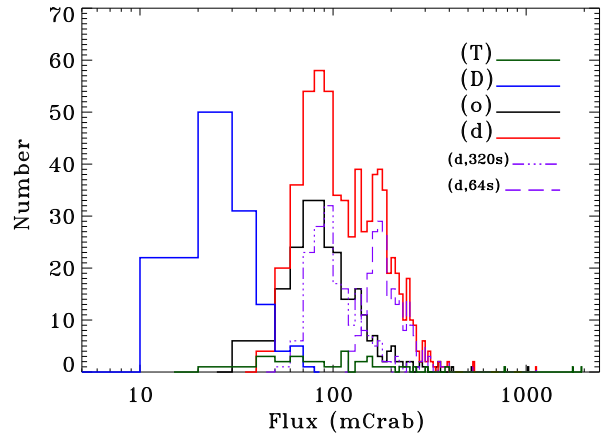}
\caption{Distributions of the flare fluxes (15--50\,keV,  in units of mCrab) 
for BAT triggers (T),
daily-averaged BATTM light curves (D),
orbital-averaged BATTM light curves (o), and 
on-board detections (d), which are 
drawn from the full catalogue. The hashed histograms represent the
on-board detections lasting 320 and  64\,s and are responsible for the two peaks 
in the (d) overall distribution at $\sim 95$ and 175\,mCrab, respectively. 
} 
\label{sfxtcat:fig:fluxes} 
\end{figure}

Figure~\ref{sfxtcat:fig:fluxes} shows the distributions of the observed 15--50\,keV 
fluxes  for the whole catalogue, depending on the detection method.
The fluxes range from $\sim 15$\,mCrab (for the daily-averages) to 
1.9\,Crab (a bright BAT trigger of IGR~J08408$-$4503) with a 
median of $\sim 105$\,mCrab. 
The medians for the four subsamples are 
134\,mCrab (T), 27\,mCrab (D), 94\,mCrab (o), and 133\,mCrab (d).  
The two peaks of the on-board (d) subsample are at$\sim 95$ and 175\,mCrab 
and are due to the detections in the 320 and 64\,s on-board images, respectively.  

\begin{figure*} 
\vspace{-4truecm}
\hspace{-1truecm}
\includegraphics[width=19.5cm]{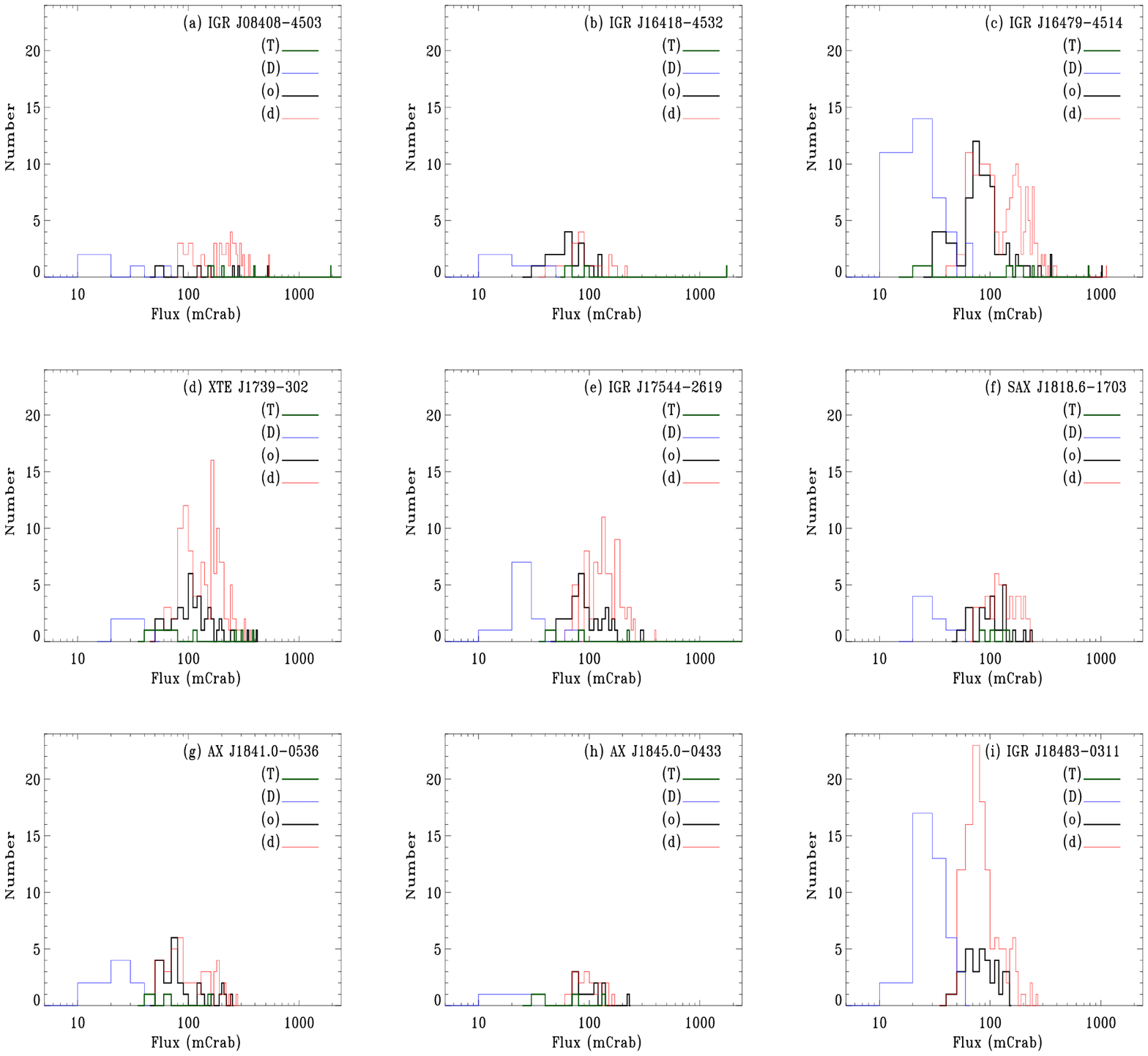}
\vspace{-6.5truecm}
\caption{Distributions of the flare fluxes (15--50\,keV) for the individual sources
for BAT triggers (T),
daily-averaged BATTM light curves (D),
orbital-averaged BATTM light curves (o), and 
on-board detections (d) in units of mCrab, drawn from the full catalogue, ordered by right ascension.
} 
\label{sfxtcat:fig:fluxes2} 
\end{figure*}

Figure~\ref{sfxtcat:fig:fluxes2} shows the flux distributions for individual sources. 
We note that we can determine the prevailing timescale for on-board  detection
when a sufficiently high number of on-board  flares is available, as in the case of 
IGR~J16479$-$4514, XTE~J1739$-$302, and IGR~J18483$-$0311.  
While the two peaks in the flux distributions 
(corresponding to  on-board detections in the 320 and 64\,s timescales) 
are equivalent for IGR~J16479$-$4514 (Figure~\ref{sfxtcat:fig:fluxes2}c), 
there is a marked preference for the 64\,s peak in 
XTE~J1739$-$302 (Figure~\ref{sfxtcat:fig:fluxes2}d) 
and  for the 320\,s peak in IGR~J18483$-$0311 (Figure~\ref{sfxtcat:fig:fluxes2}i). This means
that the on-board data imply a flare length $\ga 64$\,s for XTE~J1739$-$302
and a flare length $\ga 320$\,s for the other two. 

The flux distributions imply that the population of about a thousand SFXTs flares we observed 
is characterized by short (a few hundred seconds) and relatively bright 
(in excess of 100\,mCrab, 15--50\,keV) events. We note that these flares generally 
last less than a day in the hard X--ray, 
as demonstrated by the lower fluxes measured 
in the BATTM daily averages.  
As we have shown \citep[e.g.][and references therein]{Romano2007,Romano2011:sfxts_paperVII}, 
in the soft X--ray the picture is radically different, as the higher sensitivity of the 
focussing instrumentation allows us not only to detect the bright flares but also to follow 
the whole outburst, lasting up to several days, depending on the source. 
Clustering of X--ray flares, however, can also be used to indirectly measure the 
length of an outburst, even though the low-level emission is not detected, as we shall see 
in the following sections.

 	 \subsection{Orbital distribution of flares} \label{sfxtcat:batdistros}

The orbit-averaged BATTM light curves were screened 
to exclude bad quality points (quality flag $>0$) and referred to the solar system 
barycentre (SSB) by using the {\sc ftools} task {\sc earth2sun}.  

For the  sources with no known orbital period (IGR~J08408$-$4503 and AX~J1841.0$-$0536), 
we searched for periodicities in the 
orbit-averaged BATTM light curves within 2--1000\,d. 
Since the light curves were non-uniformly 
spaced (due to the gaps when the source was Sun-constrained), 
we applied the fast Lomb-Scargle periodogram 
technique \citep[][]{Press1989,Scargle1982,Lomb1976}. 
For each light curve, we estimated the number of independent frequencies in the 
periodograms using eq.~13 of \citet{Horne1986}. 
No statistically significant periodicities were detected. 

For those sources for which an orbital period is already known (9/11),  
we performed a standard folding analysis on the BATTM orbit-averaged light curves. 
Table~\ref{sfxtcat:tab:new_periods} shows the periods found from our analysis 
for IGR~J16465$-$4507, IGR~J16479$-$4514, IGR~J17544$-$2619, SAX~J1818.6$-$1703, 
and IGR~J18483$-$0311, where the uncertainties quoted are the 
errors derived by a Gaussian fit of the peak centroid. 

The ones found for the remainder of the sample (4/9) were consistent with the values 
reported in Table~\ref{sfxtcat:tab:datasprops}, so those values were adopted, instead. 

For all sources with a known period, we also considered the full catalogue of flares 
(Sect.~\ref{sfxtcat:batdetections}),  and we folded them with the same parameters 
as the BATTM light curves. 

\begin{figure*} 
\vspace{-4truecm}
\hspace{-1truecm}
\includegraphics[width=19.5cm]{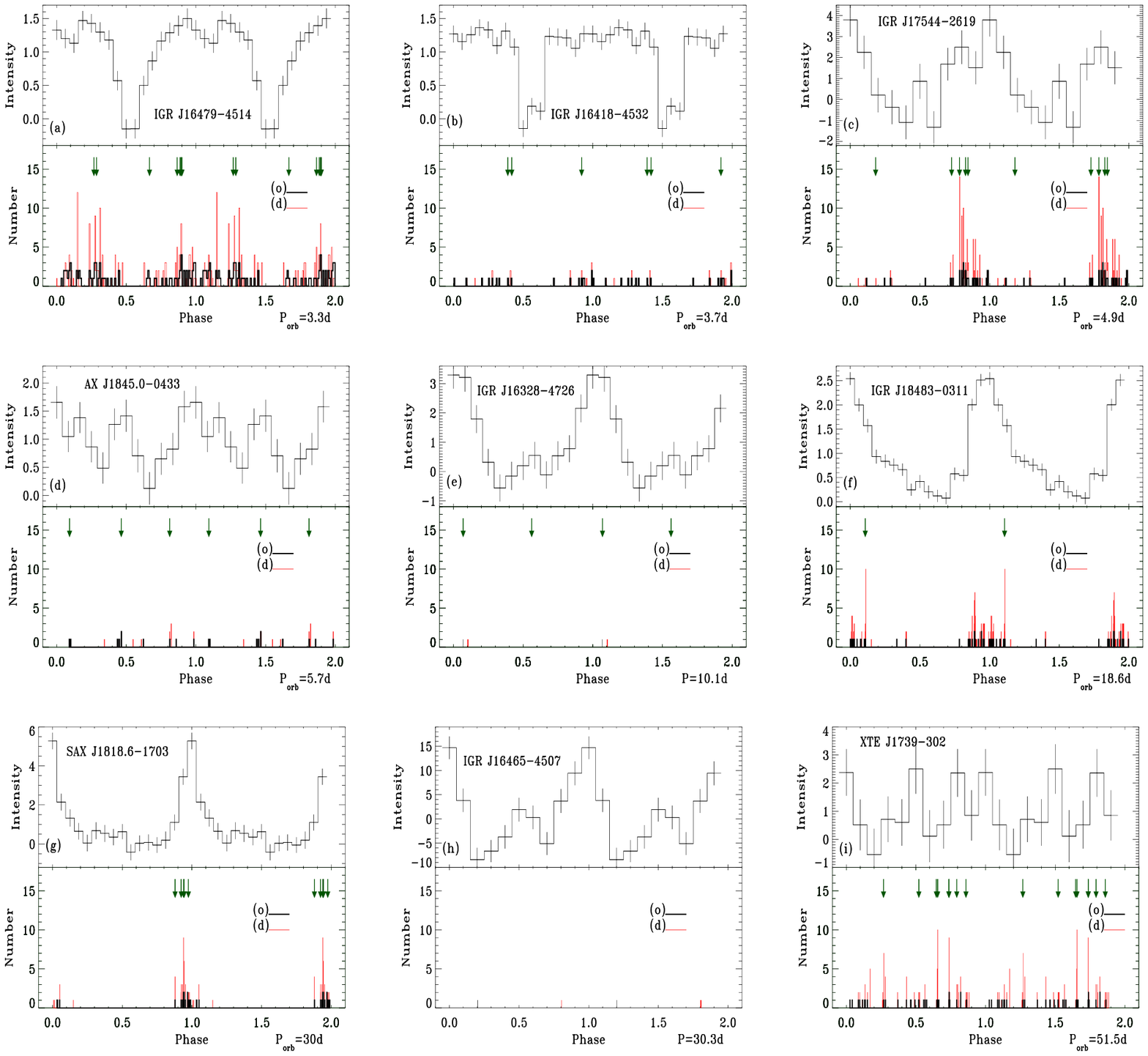}
\vspace{-6.5truecm}
\caption{{\it Top Panels}: BAT TM light curves (15--50\,keV) of the SFXT sample folded at the periods
 reported in Table~\ref{sfxtcat:tab:new_periods}, 
ordered by orbital period (reported in right lower corner of each panel).
{\it Bottom panels}: BAT on-board (d, red histograms) and 
BATTM orbital detections (o, black histograms) folded at the same periods.  
The green downward pointing arrows are the BAT triggers (T). 
} 
\label{sfxtcat:fig:folding} 
\end{figure*}

Figure~\ref{sfxtcat:fig:folding} (top panels) shows the BATTM orbital-averaged light 
curves folded at the periods reported in Table~\ref{sfxtcat:tab:new_periods}. 
The distributions of flares along the orbital period are plotted in the bottom panels of
Fig.~\ref{sfxtcat:fig:folding} with a black line marking detections in the orbit-averaged
light curves (o), a red line marking on-board detections (d), and green downward pointing arrows
marking the BAT triggers (T). The histogram bins correspond for each source and 
for each kind of datum to the maximum observed exposure during which each detection was achieved.  

\begin{figure*} 
\vspace{-4truecm}
\hspace{-1truecm}
\includegraphics[width=19.5cm]{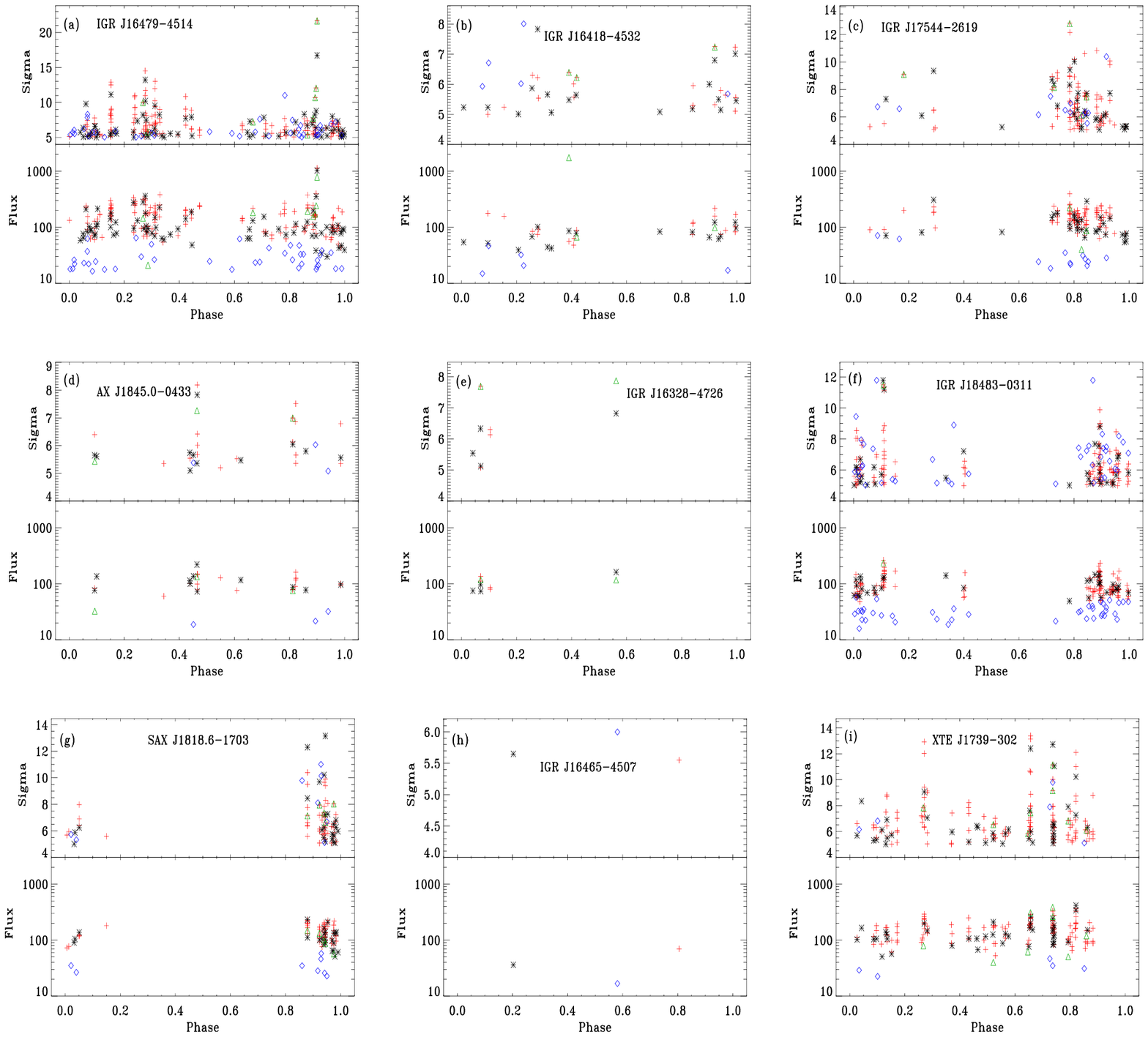}
\vspace{-6.5truecm}
\caption{{\it Top Panels}: Flare significance (in units of $\sigma$) as a function 
of orbital phase (periods in Table~\ref{sfxtcat:tab:new_periods}). 
Green triangles mark the BAT triggers (T), 
blue diamonds the BATTM daily detections (D), 
black stars the BATTM orbital detections (o),  
red crosses the  on-board detections (d). 
{\it Bottom panels}: Flare flux (in units of mCrab) as a function 
of orbital phase. The typical relative error on flux is $\sim 15$\,\%. 
} 
\label{sfxtcat:fig:folding2} 
\end{figure*}

Figure~\ref{sfxtcat:fig:folding2}  shows, as a function of orbital phase, the significance of 
each detection (in units of $\sigma$) and the flux of the detections (in units of mCrab).

\setcounter{table}{4} 
 \begin{table}
 \tabcolsep 4pt   	
 \caption{BAT orbital periods adopted for Fig~\ref{sfxtcat:fig:folding} and ~\ref{sfxtcat:fig:folding2}.\label{sfxtcat:tab:new_periods} }
 \centering
 \begin{tabular}{lcl}
\hline 
\hline
\noalign{\smallskip} 
 Name              &$P_{\rm orb}$\tablefootmark{a}        &  Epoch       \\  
                        & (d)                     &    (MJD)                             \\ 
 \noalign{\smallskip} 
 \hline 
\noalign{\smallskip}
IGR~J16328$-$4726        & $10.076$\tablefootmark{b} $\pm  0.065$         &  54256.08\tablefootmark{b} \\ 
IGR~J16418$-$4532        & $3.73886$\tablefootmark{c}  $\pm 0.00003$    &  53560.2\tablefootmark{c}   \\ 
IGR~J16465$-$4507        & $30.258  \pm  0.009$                                         &  54172.4236\tablefootmark{d}    \\  
IGR~J16479$-$4514        & $3.31935 \pm 0.00003$                               &  54170.20\tablefootmark{d}       \\  
XTE~J1739$-$302	        & $51.47$\tablefootmark{e}  $\pm0.02$                &  52698.2\tablefootmark{e}   \\ 
IGR~J17544$-$2619        & $4.93   \pm  0.07$             &  52702.9\tablefootmark{d}  \\ 
SAX~J1818.6$-$1703      & $29.99 \pm  0.08$            &  52712.3\tablefootmark{d}  \\ 
AX~J1845.0$-$0433        & $5.67$\tablefootmark{f} $\pm0.06$                 &  52708.4397\tablefootmark{f}  \\  
IGR~J18483$-$0311        & $18.56 \pm  0.07$            &  52770.6\tablefootmark{d}  \\ 
  \noalign{\smallskip}
  \hline
  \end{tabular}
\tablefoot{
\tablefoottext{a}{$P_{\rm orb}$ derived from standard folding analysis unless otherwise 
specified; errors on $P_{\rm orb}$ estimated from a Gaussian fit of the peak centroid.}
     \tablefootmark{b}{\citet{Corbet2010:16328-4726}.}
     \tablefoottext{c}{\citet{Levine2011}.}
\tablefoottext{d}{Epoch of X-ray maximum.}
    \tablefoottext{e}{\citet{Drave2010:17391_3021_period}.}
    \tablefoottext{f}{\citet{Goossens2013:18450_period}.}
}
\end{table}

         \subsection{Results for individual objects \label{sfxtcat:obj2:notes}} 

The least active sources in our sample are IGR~J16465$-$4507 and  
IGR~J16328$-$4726, as shown in Table~\ref{sfxtcat:tab:alltriggers}. 
With their orbital periods of $\sim30$ and $\sim 10$\,d, only 30 and 169 orbital cycles 
were observed, respectively and, as Fig.~\ref{sfxtcat:fig:folding}h,e and \ref{sfxtcat:fig:folding2}h,e show, 
no conclusions can be drawn on the distribution of the few flares as a function of orbital phase. 

The analysis of the data on AX~J1845.0$-$0433 (Fig.~\ref{sfxtcat:fig:folding}d and \ref{sfxtcat:fig:folding2}d)
that was followed for 463 cycles yields no marked preference with phase for the flare 
distributions and the BAT triggers,  
which occur at phase 0.09, 0.46, and 0.81. 

For IGR~J08408$-$4503 no orbital period is firmly known, despite an initial hint of a periodicity in the 
outbursts \citep{Romano2009:sfxts_paper08408} at about 35\,d with possible sub-periodicities at 11 and 24\,d.
With the current dataset this trend cannot be verified as the latter flares deviate from this pattern. 
By examining the distribution of flares in time, we note however that 
they do tend to cluster in outbursts with episodes of prolonged duration up to a day 
(see Col.~5 of Table~\ref{sfxtcat:tab:alltriggers}).  
This is consistent with a multiple-flare structure of each outburst, as observed first 
in 2008 \citep{Romano2009:sfxts_paper08408}. 

The orbital period of AX~J1841.0$-$0536 is similarly unknown. The four BAT triggers for this object 
are 383.9, 356.2, and 39.2\,d apart, and several flares cluster around them. 
The clustering is not as strong as in IGR~J08408$-$4503, so that there are detections
well away from the BAT triggers. 
We also know that this source is very active in the soft X--ray band, and indeed 
it has a very low {\it inactivity} duty cycle (IDC)\footnote{The {\it inactivity} 
duty cycle is defined in \citet{Romano2009:sfxts_paperV} for a sample of four SFXTs 
as the time each source spends {\it undetected} down to a flux limit 
of 1--3$\times10^{-12}$ erg cm$^{-2}$ s$^{-1}$, depending on the source.} 
of 28\,\%, as shown in \citet[][Fig.~1d and Table~2]{Romano2009:sfxts_paperV}. 
Given the currently scanty information, one can only speculate that an orbital period of about 3\,d,
which is comparable to that of IGR~J16479$-$4514 
(which has an inactivity duty cycle of 19\,\%, \citealt{Romano2011:sfxts_paperVI}), 
would not only be consistent with all triggers and the day-long activity around each trigger 
but would also be consistent with the relatively high soft X-ray duty cycle, as appropriate for a tight, almost circular, 
orbit around the supergiant companion. 

The transient IGR~J16479$-$4514 is by far the most active source in our sample (see Table~\ref{sfxtcat:tab:alltriggers}).
During the 891 cycles it was observed with the BAT, 
the source is detected at all orbital phases 
(Fig.~\ref{sfxtcat:fig:folding}a and \ref{sfxtcat:fig:folding2}a)
at both the daily-average and orbital-average level, except during the eclipse ($\phi=0.5$--0.6). 
This distribution closely matches that of the BATTM folded light curve.  
Furthermore, the BAT triggers cluster around the extremely narrow ranges of 
$\phi=0.88$--0.90 (5/8, including the most significant in terms of $\sigma$) 
and 0.26--0.29 (2/9), which closely match the detailed peaks of the 
BATTM folded light curve, 
while one trigger occurs at  $\phi=0.67$ (Fig.~\ref{sfxtcat:fig:folding2}a).
The picture we can draw from this information, combined with the high
activity duty cycle in the soft X--ray \citep{Romano2011:sfxts_paperVI}, 
is that of an orbit with mild eccentricity well within the accretion region 
that is determined by the wind of the supergiant companion. 

The source IGR~J16418$-$4532 
(Fig.~\ref{sfxtcat:fig:folding}b and \ref{sfxtcat:fig:folding2}b)
has a similar behaviour as IGR~J16479$-$4514, which is not surprising, given the 
very similar orbital periods, 
with flares at all phases that match the BATTM folded light curve (794 cycles),
including the deep eclipse at $\phi=0.5$--0.6. 
The BAT triggers concentrate in phase around about 0.4 and 0.9.

The source XTE~J1739$-$302  (Fig.~\ref{sfxtcat:fig:folding}i and \ref{sfxtcat:fig:folding2}i)
was only followed for 58 cycles, given the long orbital period. 
It is as active as IGR~J16479$-$4514 (see Table~\ref{sfxtcat:tab:alltriggers}) and shows 
flares (from daily- and orbital averaged and on-board data) and BAT triggers observed at all orbital phases.

The three remaining sources, IGR~J17544$-$2619, SAX~J1818.6$-$1703, and
IGR~J18483$-$0311, show a conspicuous clustering of the flares at specific phases. 
The prototypical SFXT IGR~J17544$-$2619 (Fig.~\ref{sfxtcat:fig:folding}c and \ref{sfxtcat:fig:folding2}c)
shows distributions of 
daily-,  orbital-averaged, and  on-board detections that closely match
the one of the BATTM folded light curve (constructed from 504 cycles).  
In particular, there is a clustering of flares at $\phi=0.7$--1, which  includes 
the most significant ones 
 (Fig.~\ref{sfxtcat:fig:folding2}c) and  corresponds to 
$\sim 1.5$\,d around what is probably the periastron. 
This is consistent with the findings of \citet{Clark2009:17544-2619period}. 
The BAT triggers also follow the same distribution, mainly occurring at periastron
(with one exception).  

The transient SAX~J1818.6$-$1703  (Fig.~\ref{sfxtcat:fig:folding}g and \ref{sfxtcat:fig:folding2}g)
is also characterized by a distribution of flares closely matching the BATTM
light curve (84 cycles), which includes the clustering of flares at $\phi=0.85$--1.05
 corresponding to $\sim 6$\,d. This is comparable 
to the width of the peak in the BATTM folded light curve and consistent with the findings 
of \citet{Bird2009:sax1818.6_period}, 
who observed that SAX~J1818.6$-$1703 shows a duty cycle of 4--6\,d during a 30\,d period. 

The source IGR~J18483$-$0311  (Fig.~\ref{sfxtcat:fig:folding}f and \ref{sfxtcat:fig:folding2}f) is 
generally considered an SFXT with an intermediate dynamical range and 
also shows a tight match between the distribution of flares and the BATTM light curves (162 cycles). 
We observe a strong clustering in phase of flares at $\phi=0.87$--1.1, which corresponds to 
$\sim 4.3$\,d. This is comparable to the width of the peak in the BATTM folded light curve, 
presumably the periastron. 
This is consistent with the structure of the folded light curve observed in the soft X--ray \citep{Romano2010:sfxts_18483}
and is indicative of a large orbit with some eccentricity, as derived by \citet{Romano2010:sfxts_18483}, 
who find $e=0.4$ by modelling the soft X--ray light curve with the \citet{Ducci2009} clumpy wind model.

 	 \section{SFXT perspectives for future missions} \label{sfxtcat:perspectives}

\begin{figure*} 
\vspace{-3truecm}
\hspace{-1truecm}
\includegraphics[width=19.5cm,height=18cm]{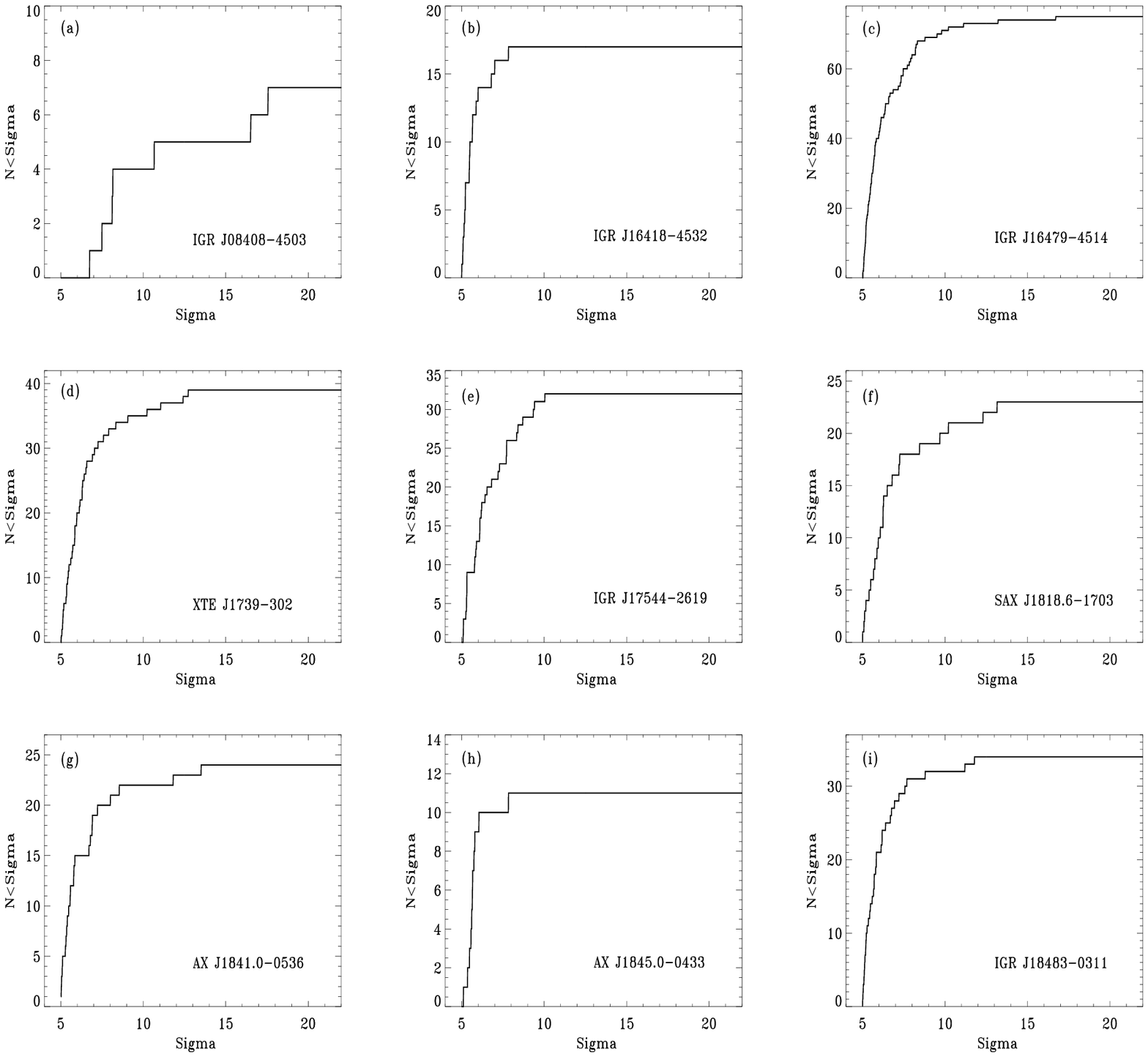}
\vspace{-4truecm}
\caption{Cumulative distributions of $\sigma$ for 
BATTM orbital detections (corresponding to (o) in Figs.~\ref{sfxtcat:fig:folding} and \ref{sfxtcat:fig:folding2})   
for our 100 months of data collected with BAT.  We note that 
$5\sigma$ detections for an average orbit correspond to 
$1.46\times10^{-9}$ erg cm$^{-2}$ s$^{-1}$ in the 15--150\,keV band 
($8.24\times10^{-10}$ erg cm$^{-2}$ s$^{-1}$, 15--50\,keV). 
We also note that the typical duration for an average orbit is 800\,s for our sample. 
} 
\label{sfxtcat:fig:sigmas} 
\end{figure*} 

As an interesting application of this work, our catalogue can be used to estimate the outcomes of and plan 
observations of SFXTs from future hard X--ray observatories, in particular for those carrying a wide 
field of view (FOV) monitor.  

We consider the fraction of days on which the source was in the BAT FOV as a good proxy for observability.  
Indeed, for each source, BAT made at least some observations on a fraction of days that varies 
from source to source but is typically $\sim 90$\,\%.   
The primary cause of long observational gaps were  Sun constraints on the pointing direction of \sw, 
so we call this overall fraction the `seasonal visibility'. 
In Table~\ref{sfxtcat:tab:expected_flares} we report the expected number of flares for 
the sources in our sample for a 5-year baseline. 
This number was corrected for the seasonal visibility. 
Preliminary results of our work were indeed used for LOFT \citep{FerociLOFT_long}  
simulations \citep{Bozzo2013:COSPAR_sfxt}. 
We note that the accuracy of the expected number of flares
is strongly dependent on the actual number of observed flares it is based upon. 

Furthermore, Fig.~\ref{sfxtcat:fig:sigmas} shows the cumulative distributions of $\sigma$ for the 
BATTM orbital detections. 
By considering that 
$5\sigma$ detections for an average orbit  
correspond to 
$1.46\times10^{-9}$ erg cm$^{-2}$ s$^{-1}$ in the 15--150\,keV band 
($8.24\times10^{-10}$ erg cm$^{-2}$ s$^{-1}$, 15--50\,keV), the individual plots can be used to 
predict the number of flares for a given limiting flux. 
For instance, let us estimate the number of flares in excess of 100\,mCrab (15--50\,keV). 
This flux corresponds to 7.75\,$\sigma$ in the 15--50\,keV band at the 
orbital-averaged level (see Sect.~\ref{sfxtcat:batdetections}) or a 
flux of $2.3\times10^{-9}$ erg cm$^{-2}$ s$^{-1}$ in the 15--150\,keV band 
($4.0\times10^{-9}$ erg cm$^{-2}$ s$^{-1}$, 15--50\,keV). 
From Fig.~\ref{sfxtcat:fig:sigmas}c we derive for IGR~J16479$-$4514  that
61 flares  are below 7.75\,$\sigma$ and  that $75-61=14$ flares are above 7.75\,$\sigma$. 
After correction for the seasonal visibility, we thus obtain 10 flares during a 5-year mission.
By adopting the same procedure for the whole sample, in a 5-year mission 
we expect a total of 
32 flares in excess of 100\,mCrab. 
Similarly, for a 5-year mission, we obtain 
48 flares in excess of 90\,mCrab (6.98\,$\sigma$),
73 flares in excess of 80\,mCrab (6.20\,$\sigma$), and 
130 flares in excess of 70\,mCrab (5.43\,$\sigma$). 

\setcounter{table}{5} 
 \begin{table}
 \tabcolsep 4pt   	
 \caption{Expected number of flares 
in excess of $1.46\times10^{-9}$ erg cm$^{-2}$ s$^{-1}$ (15--150\,keV band) 
for a 5-year mission from the SFXT sample.
\label{sfxtcat:tab:expected_flares} }
 \centering
 \begin{tabular}{lrr}
\hline 
\hline
\noalign{\smallskip} 
Name                                                    & Seasonal                                 &Number of             \\  
                                                             &  visibility\tablefootmark{a}      & Flares   (o)\tablefootmark{b}                  \\ 
 \noalign{\smallskip} 
 \hline 
\noalign{\smallskip}
IGR~J08408$-$4503                            &    0.90          &4 \\ 
IGR~J16328$-$4726                  &  0.90         &2  \\ 
IGR~J16418$-$4532                   &  0.88         &11 \\ 
 IGR~J16465$-$4507                            &    0.91           &1  \\ 
IGR~J16479$-$4514                            &   0.87            &51  \\ 
XTE~J1739$-$302	                            &   0.87            &27  \\ 
IGR~J17544$-$2619                            &   0.85            & 22\\ 
SAX~J1818.6$-$1703                          &   0.86             &16 \\ 
AX~J1841.0$-$0536                            &   0.87             & 17  \\ 
AX~J1845.0$-$0433                            &   0.90             &7   \\ 
IGR~J18483$-$0311                            &  0.90              &23 \\ 
  \noalign{\smallskip}
  \hline
  \noalign{\smallskip}
  Totals                                              &   &   185  \\ 
  \noalign{\smallskip}
  \hline
  \end{tabular}
\tablefoot{
\tablefoottext{a}{Fraction of year during which each source was observed because of several visibility constraints, 
including Sun constraints.}
\tablefoottext{b}{Uncertainties are of the order of the Poisson error on the quoted number.}
}
\end{table}

We note that our expectations are conservative lower limits, as they are based
on the currently known population of SFXTs, which is bound to increase as the surveys of the
Galactic plane become deeper and outbursts of new SFXTs are observed.

 	 \section{Discussion} \label{sfxtcat:discussion}

%
The mechanisms responsible for the bright short outbursts in SFXTs  
are still not known well. The most credited models can be roughly divided in 
two categories: 
{\it i)} models for which the X-ray variability exclusively
depends on the properties of the geometry and inhomogeneity
of the stellar wind from the donor star, and 
{\it ii)} models in which the accretion mechanisms which link the observed high 
dynamic ranges to the properties of the compact object can be explained by 
assuming only modest inhomogeneities in the density and/or velocity 
in the supergiant  wind. 

In the spherically symmetric clumpy wind model, the short flares are produced by 
accretion of massive clumps (10$^{22}$--10$^{23}$\,g) in the supergiant winds 
\citep[e.g.][]{zand2005,Negueruela2008,Walter2007}, 
which are believed to be strongly inhomogeneous \citep[e.g.][]{Oskinova2007}
with large density contrasts (10$^{4}$--10$^{5}$). 
In this model, SFXTs should have generally wider orbits than persistent HMXBs
\citep[][]{Negueruela2008,Chaty2008:obscured}. 
In particular, a key prediction is strong variations in the column density, 
which have recently been observed in  IGR~J17544$-$2619 \citep{Rampy2009:suzaku17544},
IGR~J08408$-$4503 \citep{Romano2009:sfxts_paper08408} and, most spectacularly, 
in AX~J1841.0$-$0536 \citep{Bozzo2011:18410}. 
For \citet{Sidoli2007}, the outbursts can be due to 
the presence of an equatorial wind component, denser, possibly clumpy, 
and slower than the symmetric polar wind from the blue supergiant, which is inclined with
respect to the orbital plane of the system. This could be the case of 
IGR~J11215--5952 \citep{Sidoli2007}. 
\citet{Ducci2009}  developed a more detailed clumpy stellar wind model 
(considering both spherical and non-spherical geometry) for OB supergiants in HMXBs, 
which assumes that a fraction of the wind is in clumps with power-law mass and size distributions.
This model allows a direct comparison to the X-ray properties derived from the light 
curves of both SFXTs and persistent HMXBs and was successfully applied to 
IGR~J08408$-$4503, IGR~J18483$-$0311, and IGR~J16418$-$4532 
\citep[][respectively]{Romano2009:sfxts_paper08408,Romano2010:sfxts_18483,Romano2012:sfxts_16418}. 

Alternatively, the high dynamic range in SFXTs can be explained 
without invoking large variations in the density and/or velocity of the supergiant wind
in terms of gated mechanisms \citep{Stella1986,Grebenev2007,Bozzo2008}, 
due to which the accretion flow is halted by a magnetic
or a centrifugal barrier, dependent on the properties of the NS, 
its $P_{\rm spin}$, and the strength of its magnetic field $B$. 
In particular, as this model works well for the SFXT prototype
IGR~J17544$-$2619,  \citet{Bozzo2008} conclude that 
the high observed dynamic range can be produced in SFXTs 
provided they host neutron stars with long  $P_{\rm spin}\ga 1000$\,s 
and magnetar-like $B\ga10^{14}$\,G fields.

\begin{figure*} 
\vspace{-0.3truecm}
\includegraphics[width=9cm]{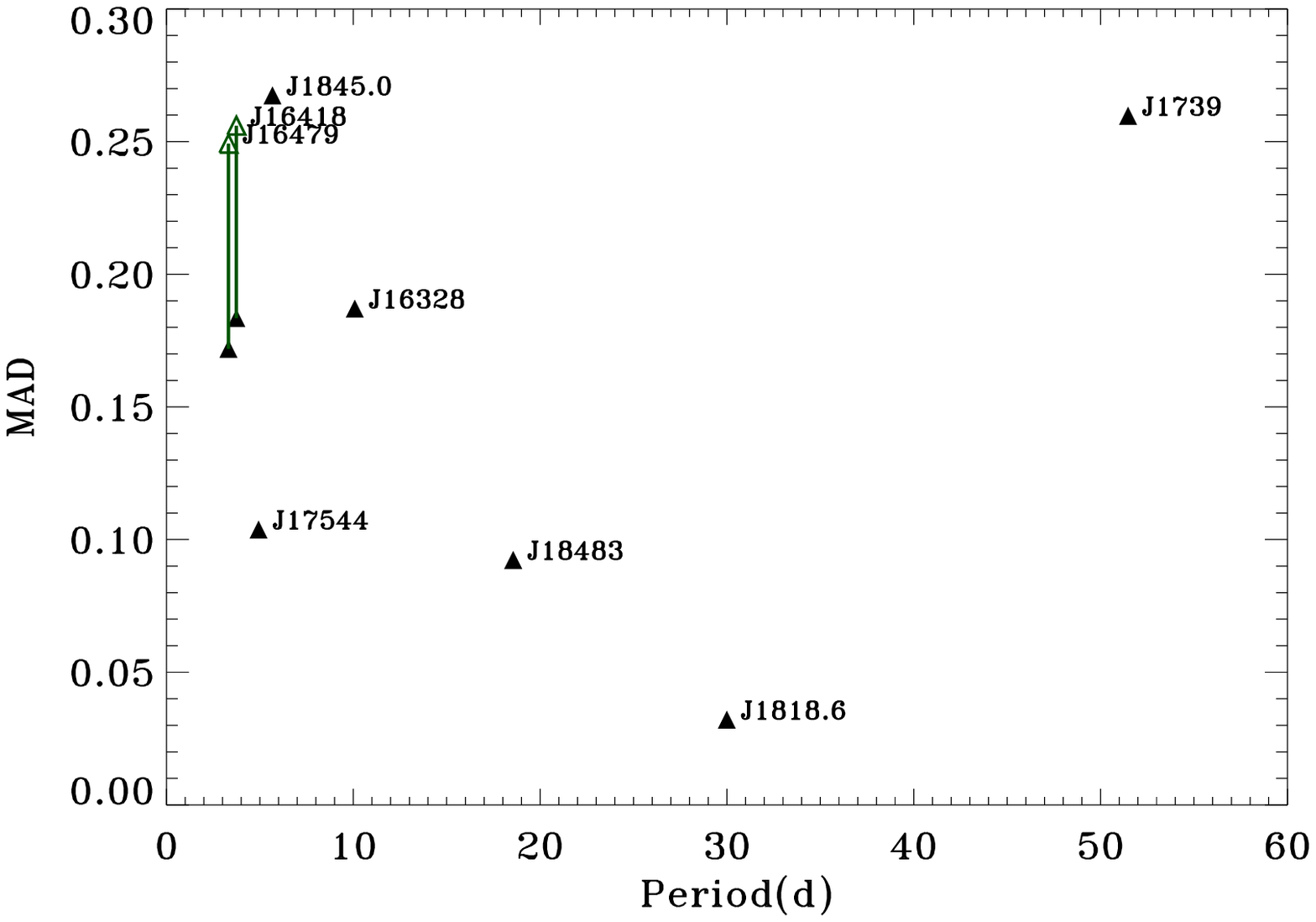}
\includegraphics[width=9cm]{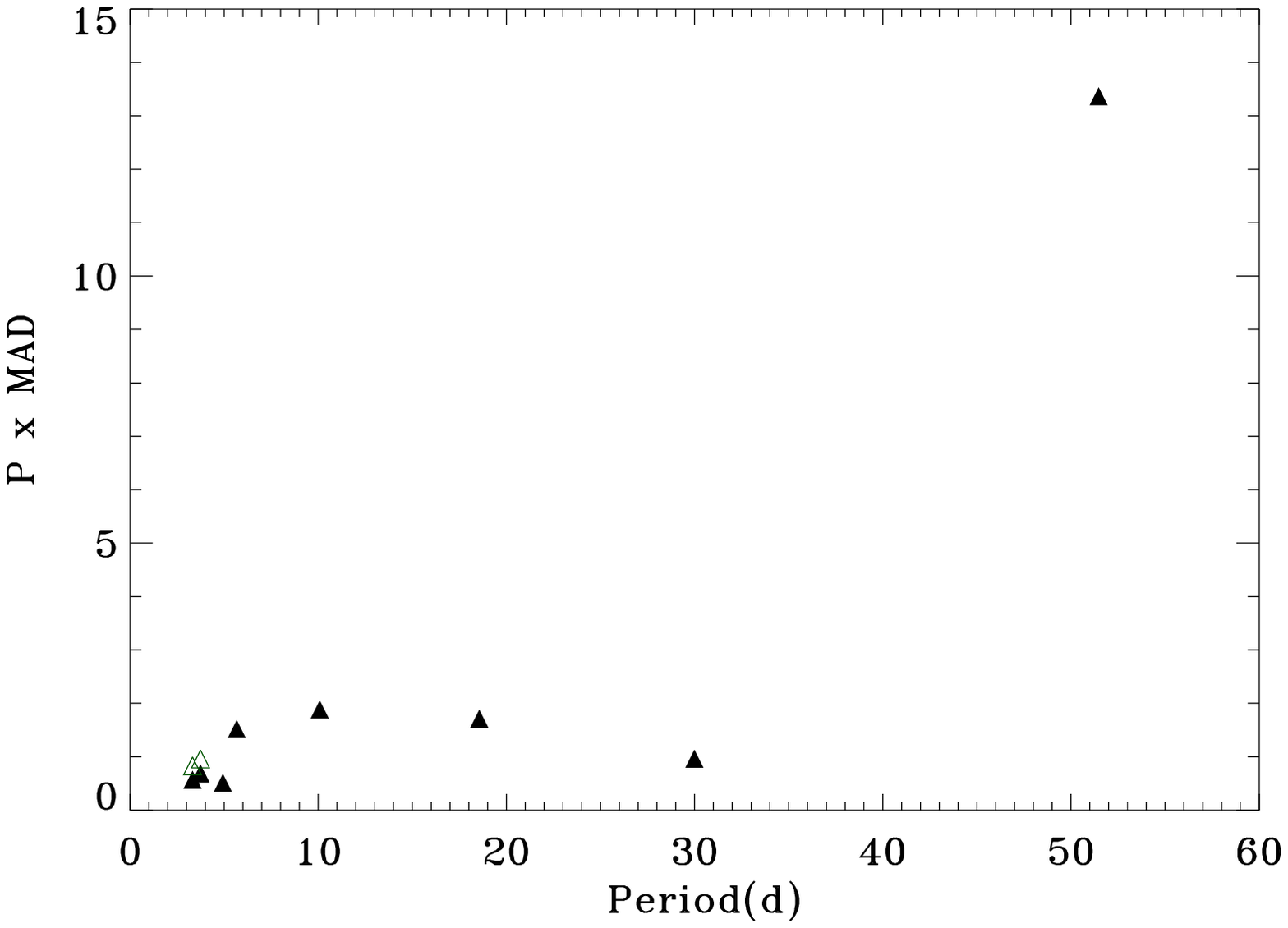}
\caption{{\it Left Panel}: Mean absolute deviation of the flare phases as a function of orbital period. 
{\it Right Panel}: Mean absolute deviation of the flare phases multiplied by the orbital period as a 
function of orbital period. The dataset is the orbital detections (o, filled black points). 
The empty green points are obtained by simulating a population of flares for 
IGR~J16479$-$4514 and IGR~J16418$-$4532 unaffected by the presence of the eclipse. 
} 
\label{sfxtcat:fig:MAD} 
\end{figure*} 

In this paper,  we have examined over a thousand detections from 11 SFXTs, 
the great majority of confirmed SFXTs, 
as observed by \sw/BAT in the hard X--ray during the first 100 months of its mission. 
Since we have applied a 
5-$\sigma$ cut to the available BATTM and on-board detections, 
our catalog is a flux limited sample of flares, which is complete down to 15--150\,keV fluxes of  
$\sim6\times10^{-10}$ erg cm$^{-2}$ s$^{-1}$  (daily timescale) and  
$\sim 1.5\times10^{-9}$ erg cm$^{-2}$ s$^{-1}$ (orbital timescale, $\sim 800$\,s). 
For the first time, therefore, it is possible to consider this homogeneous set of 
data to try to address long standing issues related to the nature 
of their emission. 

Our sample of flares shows marked differences in behaviour among individual objects 
in terms of both overall activity, (i.e., number of flares observed,  their intensity and duration, 
Table~\ref{sfxtcat:tab:alltriggers},  Figs.~\ref{sfxtcat:fig:fluxes2} and \ref{sfxtcat:fig:durations}) 
and clustering of the flares in phase. 
The distribution of flare fluxes (Fig.~\ref{sfxtcat:fig:fluxes2}) in the 15--50\,keV energy band 
show that they range from 
$\sim 15$\,mCrab to about 2\,Crab ($1.9\times10^{-10}$--$2.5\times10^{-8}$  erg cm$^{-2}$ s$^{-1}$), 
which corresponds to luminosities of $L_{\rm 15-50\,keV} \sim 6\times10^{35}$ erg s$^{-1}$   to  
$L_{\rm 15-50\,keV}  \sim 7\times10^{37}$ erg s$^{-1}$ 
(assuming a common distance of 5 kpc, which is the average of the sample in 
Table~\ref{sfxtcat:tab:datapositions}). 
The median flux is $\sim 105$\,mCrab, which corresponds to 
 $L_{\rm 15-50\,keV}  \sim 4\times10^{36}$  erg s$^{-1}$. 
In the clumpy stellar wind scenario, flares with luminosities of
$10^{36}$--$10^{37}$ erg s$^{-1}$ and durations of hours can be produced by
the accretion of clumps of wind with mass of $10^{22-23}$\, g \citep[][]{Walter2007}.

The distribution of the flare durations (Fig.~\ref{sfxtcat:fig:durations}
and  \ref{sfxtcat:fig:example_d}) shows that they range from a few minutes up 
to a few hours. 
In the clumpy wind models, luminosity variations can occur on timescales 
of the order of hours, the time required for the accretion of individual clumps,
or the crossing time of the NS through the equatorial plane (in the model proposed by 
\citealt{Sidoli2007}). 
In the gating mechanisms scenario, transitions across the NS regimes
are caused by small variations in the wind local density \citep[][]{Grebenev2007,Bozzo2008}, 
which also occur on timescales of the order of hours. 
Therefore, the durations of the BAT flares do not allow us to discriminate 
between the two scenarios.

Another issue we can tackle is the impact of the binary geometry on the overall 
hard X-ray emission.  
Figure~\ref{sfxtcat:fig:folding} shows the distribution of flares with the orbital phase 
with the objects ordered by length of the orbital period $P_{\rm orb}$. 
As stated in detail in Sect.~\ref{sfxtcat:batdistros}, little can be inferred for 
IGR~J16465$-$4507 and IGR~J16328$-$4726 due to the low number of flares. 
However, we note a definite trend in 
the clustering of flares for the remainder of the objects. 
Sources with very short periods (IGR~J16479$-$4514 and IGR~J16418$-$4532, 
$P_{\rm orb}\sim 3$\,d) are detected at all orbital phases at both the daily-average 
and orbital-average level, except during their eclipse (at $\phi=0.5$--0.6). 
In combination with the observed soft X--ray properties of a high activity duty cycle 
\citep{Romano2011:sfxts_paperVI,Romano2012:sfxts_16418}, 
this is clearly consistent with a tight circular or mildly eccentric orbit, well within the accretion 
region determined by the wind of the supergiant companion. 
As $P_{\rm orb}$ increases, we note that the clustering of flares 
with phase becomes more and more pronounced, as can be best seen in the 
case of IGR~J18483$-$0311 and SAX~J1818.6$-$1703 
($P_{\rm orb}\sim18$ and 30\,d, respectively), which can be 
readily explained in terms of larger and probably more eccentric orbits. 
The only exception to this trend is XTE~J1739$-$302, whose properties
of a large number of flares observed at all phases, and a high soft X-ray 
activity duty cycle \citep{Romano2011:sfxts_paperVI}, 
would be more similar to those of a short period binary (such as IGR~J16479$-$4514)
than to a binary with a long $P_{\rm orb}\sim51$\,d. 
Similarly, if we considered AX~J1841.0$-$0536, whose orbital period is still unknown,  
we could speculate that an orbital period of about 3\,d (comparable to that of IGR~J16479$-$4514) 
would not only be consistent with all observed BAT triggers and the day-long activity around each trigger 
but also with the relatively high soft X-ray duty cycle \citep{Romano2009:sfxts_paperV}, 
which is typical of a tight, almost circular, orbit around the supergiant companion. 

To better quantify these statements, we calculated the mean absolute deviation 
(MAD)\footnote{ $$\mathrm{Mean~Absolute~Deviation}=\frac{1}{N-1}\sum_{j=1}^{N}|x_j-\overline{x}|,$$ 
where $\overline{x}$ is the mean of $x=(x_1,...,x_N)$.   }
of the flare phases for each source with known 
orbital period. 
The MAD multiplied by the orbital period can be considered a good proxy 
for the half-length of the duty-cycle. 
Figure~\ref{sfxtcat:fig:MAD} (left panel) shows the MAD, where the black points 
are derived using all orbital detections, while the green ones are the 
MAD values we obtained by simulating a population of flares for 
IGR~J16479$-$4514 and IGR~J16418$-$4532 unaffected by the presence of the eclipse
(by `filling' the eclipses with flares randomly chosen among the observed distribution).
Figure~\ref{sfxtcat:fig:MAD} (right) reports the MAD multiplied by the orbital period, 
and it shows that the duty-cycle is in the order of a few days. 
In both panels of Figure~\ref{sfxtcat:fig:MAD}, XTE~J1739$-$302 is a clear outlier, 
thus suggesting that the $\sim 51$\,d period may not be the orbital period  but possibly a
superorbital period, such as were recently discovered for IGR~J16479$-$4514 and IGR~J16418$-$4532
\citep{CorbetKrimm2013:superorbital}. 

Clumpy wind and gating mechanism models can both explain
the observed increase of clustering of flares with the
eccentricity. In both scenarios, the clustering of flares
is expected at the periastron passage, where the number density
of clumps is larger (compared to the apastron region)
and the higher mass accretion rate
($\dot{M}_{\rm acc} \propto \rho v_{\rm rel}^{-3}$,
where $\rho$ is the wind density and $v_{\rm rel}$ is the relative velocity
between the NS and the wind; e.g. \citealt{Waters1989}) 
can open the centrifugal/magnetic barriers, leading to direct accretion.

 	 \section{Summary} \label{sfxtcat:conclusions}

We have presented a catalogue of over a thousand BAT flares on 11 SFXTs, 
down to 15--150\,keV fluxes of  $\sim6\times10^{-10}$ erg cm$^{-2}$ s$^{-1}$  (daily timescale) and  
$\sim 1.5\times10^{-9}$ erg cm$^{-2}$ s$^{-1}$ (orbital timescale, averaging $\sim 800$\,s) 
spanning the initial 100 months
of the \sw\ mission. We showed that these flares, 
which are for the large majority previously unpublished, 
are short (a few hundred seconds, as predicted by 
all models for SFXT outburst emission) 
and relatively bright (in excess of 100\,mCrab, 15--50\,keV) events, 
lasting much less than a day in the hard X--ray. 
The outbursts are, however, 
as shown by the clustering of flares in phase space, a much longer phenomenon,
lasting up to a few days, as previously discovered from deeper soft X--ray observations
 \citep[e.g.][and references therein]{Romano2007,Romano2011:sfxts_paperVII}. 
In particular, we observe a trend of clustering of flares at some phases as $P_{\rm orb}$ increases,  
which is  consistent with a progression from tight circular or mildly eccentric orbits at short periods,  
to wider and more eccentric orbits at longer orbital periods. 

This large dataset (the largest in a single publication) can be used to further 
probe the properties of the high and intermediate emission states in SFXTs, and 
to infer the properties of these binary systems, especially in conjunction with 
flares detected by other current or future missions. 
For the latter, we have also provided a simple recipe to estimate the number of 
flares per year each source is likely to produce as a function of 
detection threshold/limiting flux.

\begin{acknowledgements}
We thank W.\ Baumgartner and C.\ Markwardt for support with the BAT survey products,
and E.\ Bozzo and M.\ Capalbi, for helpful discussions. 
We also thank the anonymous referee for constructive comments that helped improve the paper. 
We acknowledge financial contribution from NASA contract NAS5-00136 at PSU.  
HAK acknowledges NASA Swift GO grants NNX09AU85G, 
NNX12AD32G, NNX12AE57G, and NNX13AC75G. 
This work made use of the results of the Swift/BAT hard X-ray transient monitor:  
\href{http://swift.gsfc.nasa.gov/docs/swift/results/transients/}{http://swift.gsfc.nasa.gov/docs/swift/results/transients/. }
\end{acknowledgements}

\vfill

\begin{appendix}
\section[T]{Supplementary tables}

In Tables~\ref{sfxtcat:tab:08408dets}--\ref{sfxtcat:tab:18483dets}, 
we report the results on individual sources as an abridged list of detections--we 
selected the brightest flare for each day the source was detected--as marked by an (o) and (d), 
for orbital-averaged  and  on-board detection, respectively. 
The detections from the daily-averaged BATTM light curves are marked by a (D). 
We also report the subsample of BAT triggers, marked by a (T), along with the 
trigger number (Col.~6). 

\setcounter{table}{0} 
 \begin{table} 	
 \tabcolsep 4pt         
 \caption{BAT detections of IGR~J08408$-$4503. \label{sfxtcat:tab:08408dets}} 	
 \centering
 \begin{tabular}{rrcrrl} 
 \hline 
 \hline 
 \noalign{\smallskip} 
MJD  & Date & Detection\tablefootmark{a} & Flux\tablefootmark{b}   & S/N  &Trigger  	 	          	    \\ 
        &      &  	&     (mCrab)         & 	 &  \#    	    \\ 
 \noalign{\smallskip} 
 \hline 
 \noalign{\smallskip} 
53853	&	2006-04-28	&	d	&	101	&	5.60	&		\\
53909	&	2006-06-23	&	o	&	58	&	6.74	&		\\
54012	&	2006-10-04	&	d	&	102	&	6.28	&		\\
54012	&	2006-10-04	&	T	&	 	&	8.08	&	232309	\\
54012	&	2006-10-04	&	o	&	85	&	8.11	&		\\
54214	&	2007-04-24	&	d	&	135	&	5.38	&		\\
54652	&	2008-07-05	&	d	&	198	&	7.38	&		\\
54652	&	2008-07-05	&	T	&	 152	&	7.38	&	316063	\\
54730	&	2008-09-21	&	D	&	17	&	6.85	&		\\
54730	&	2008-09-21	&	d	&	535	&	6.61	&		\\
54730	&	2008-09-21	&	T	&	391	&	10.00	&	325461	\\
54730	&	2008-09-21	&	o	&	527	&	7.49	&		\\
54968	&	2009-05-17	&	D	&	18	&	5.28	&		\\
55071	&	2009-08-28	&	D	&	31	&	5.50	&		\\
55071	&	2009-08-28	&	T	& 1930	 	&	6.62	&{\it 361128}\tablefootmark{c} 	\\
55071	&	2009-08-28	&	o	&	157	&	10.66	&		\\
55071	&	2009-08-28	&	d	&	280	&	10.30	&		\\
55071	&	2009-08-28	&	T	&	 	&	10.26	&	{\it 361129}\tablefootmark{c} 	\\
55283	&	2010-03-28	&	d	&	266	&	8.72	&		\\
55283	&	2010-03-28	&	T	&	170 	&	8.68	&	417420	\\
55283      &     2010-03-28  &o\tablefootmark{d}  &          40&  3.93  & \\   
55798	&	2011-08-25	&	D	&	69	&	15.55	&		\\
55798	&	2011-08-25	&	o	&	282	&	8.15	&		\\
55798	&	2011-08-25	&	T	&	 203	&	7.28	&	501368	\\
55798	&	2011-08-25	&	d	&	356	&	13.10	&		\\
 \noalign{\smallskip}
  \hline
   \end{tabular}
\tablefoot{
\tablefoottext{a}{Detection in the daily (D) or orbital (o) sampling; (T) BAT trigger; (d) on board detection.}
\tablefoottext{b}{Flux in mCrab the 15--50\,keV energy range.  }
\tablefoottext{c}{Double trigger. }
\tablefoottext{d}{Short (64s) flare that triggered onboard.  The BATTM monitor result here reported 
is for a 1000s interval, at a lower total significance.}
}  
\end{table}

\setcounter{table}{1} 
 \begin{table} 	
 \tabcolsep 4pt         
 \caption{BAT detections of IGR~J16328$-$4726.  \label{sfxtcat:tab:16328dets}} 	
  \centering
 \begin{tabular}{rrcrrl} 
 \hline 
 \hline 
 \noalign{\smallskip} 
MJD  & Date & Detection\tablefootmark{a} & Flux\tablefootmark{b}   & S/N  &Trigger  	 	          	    \\ 
        &      &  	&     (mCrab)         & 	 &  \#    	    \\ 
 \noalign{\smallskip} 
 \hline 
 \noalign{\smallskip} 
54226	&	2007-05-06	&	d	&	86	&	6.13	&		\\
54992	&	2009-06-10	&	o	&	96	&	6.33	&		\\
54992	&	2009-06-10	&	T	&	 118	&	7.69	&	354542	\\
54992	&	2009-06-10	&	d	&	135	&	5.08	&		\\
55254	&	2010-02-27	&	o	&	75	&	5.54	&		\\
55924	&	2011-12-29	&	T	&	 117	&	7.87	&	510701	\\
55924      &  2011-12-29       & o           &    162       & 6.82   &   \\      
  \noalign{\smallskip}
  \hline
   \end{tabular}
\tablefoot{
\tablefoottext{a}{Detection in the daily (D) or orbital (o) sampling; (T) BAT trigger; (d) on board detection.}
\tablefoottext{b}{Flux in mCrab the 15--50\,keV energy range.  }
}  
\end{table} 

\setcounter{table}{2} 
 \begin{table} 	
 \tabcolsep 4pt         
 \caption{BAT detections of IGR~J16418$-$4532. \label{sfxtcat:tab:16418dets}} 	
 \centering
 \begin{tabular}{rrcrrl} 
 \hline 
 \hline 
 \noalign{\smallskip} 
MJD  & Date & Detection\tablefootmark{a} & Flux\tablefootmark{b}   & S/N  &Trigger  	 	          	    \\ 
        &      &  	&     (mCrab)         & 	 &  \#    	    \\ 
 \noalign{\smallskip} 
 \hline 
 \noalign{\smallskip} 
53418	&	2005-02-17	&	D	&	17	&	5.68	&		\\
53418	&	2005-02-17	&	o	&	52	&	5.23	&		\\
53430	&	2005-03-01	&	o	&	44	&	5.66	&		\\
53576	&	2005-07-25	&	D	&	21	&	8.01	&		\\
53639	&	2005-09-26	&	D	&	15	&	5.93	&		\\
53639	&	2005-09-26	&	o	&	42	&	5.06	&		\\
53821	&	2006-03-27	&	o	&	97	&	5.45	&		\\
54221	&	2007-05-01	&	d	&	88	&	5.64	&		\\
54296	&	2007-07-15	&	o	&	62	&	5.50	&		\\
54495	&	2008-01-30	&	d	&	177	&	5.00	&		\\
54520	&	2008-02-24	&	o	&	83	&	5.08	&		\\
54546	&	2008-03-21	&	o	&	122	&	6.80	&		\\
54546	&	2008-03-21	&	d	&	217	&	7.25	&		\\
54546	&	2008-03-21	&	T	&	 98	&	7.23	&	307208	\\
54580	&	2008-04-24	&	o	&	66	&	6.00	&		\\
54666	&	2008-07-19	&	o	&	82	&	5.18	&		\\
54666	&	2008-07-19	&	d	&	122	&	5.95	&		\\
54742	&	2008-10-03	&	o	&	39	&	5.01	&		\\
55796	&	2011-08-23	&	o	&	54	&	5.23	&		\\
55834	&	2011-09-30	&	d	&	157	&	5.24	&		\\
55844	&	2011-10-10	&	o	&	123	&	7.01	&		\\
55844	&	2011-10-10	&	d	&	168	&	5.56	&		\\
56017	&	2012-03-31	&	D	&	46	&	6.71	&		\\
56017	&	2012-03-31	&	o	&	101	&	7.83	&		\\
56017	&	2012-03-31	&	d	&	92	&	5.54	&		\\
56081	&	2012-06-03	&	D	&	32	&	6.02	&		\\
56081	&	2012-06-03	&	o	&	78	&	5.64	&		\\
56081	&	2012-06-03	&	d	&	89	&	6.23	&		\\
56081	&	2012-06-03	&	T	&	 67	&	6.22	&	523489	\\
56113	&	2012-07-05	&	o	&	69	&	5.15	&		\\
56113	&	2012-07-05	&	d	&	75	&	5.80	&		\\
56384	&	2013-04-02	&	o	&	86	&	5.48	&		\\
56384	&	2013-04-02	&	T	&	1741 &	6.39	&	552677	\\
56384	&	2013-04-02	&	d	&	63	&	5.57	&		\\
    
 \noalign{\smallskip}
  \hline
   \end{tabular}
\tablefoot{
\tablefoottext{a}{Detection in the daily (D) or orbital (o) sampling; (T) BAT trigger; (d) on board detection.}
\tablefoottext{b}{Flux in mCrab the 15--50\,keV energy range.  }
}  
\end{table} 

\setcounter{table}{3} 
 \begin{table} 	
 \tabcolsep 4pt         
 \caption{BAT detections of IGR~J16465$-$4507.  \label{sfxtcat:tab:16465dets}} 	
 \centering
 \begin{tabular}{rrcrrl} 
 \hline 
 \hline 
 \noalign{\smallskip} 
MJD  & Date & Detection\tablefootmark{a} & Flux\tablefootmark{b}   & S/N  &Trigger  	 	          	    \\ 
        &      &  	&     (mCrab)         & 	 &  \#    	    \\ 
 \noalign{\smallskip} 
 \hline 
 \noalign{\smallskip} 
53633   &       2005-09-20      &       o       &       36      &       5.65    &               \\
54190   &       2007-03-31      &       D       &       17      &       6.00    &               \\
54529   &       2008-03-04      &       d       &       70      &       5.55    &               \\
\noalign{\smallskip}
  \hline
   \end{tabular}
\tablefoot{
\tablefoottext{a}{Detection in the daily (D) or orbital (o) sampling; (T) BAT trigger; (d) on board detection.}
\tablefoottext{b}{Flux in mCrab the 15--50\,keV energy range.  }
}  
\end{table}

\setcounter{table}{4} 
 \begin{table} 	
 \tabcolsep 4pt         
 \caption{BAT  detections of IGR~J16479$-$4514. \label{sfxtcat:tab:16479dets}} 	
 \centering
 \begin{tabular}{rrcrrl} 
 \hline 
 \hline 
 \noalign{\smallskip} 
MJD  & Date & Detection\tablefootmark{a} & Flux\tablefootmark{b}   & S/N  &Trigger  	 	          	    \\ 
        &      &  	&     (mCrab)         & 	 &  \#    	    \\ 
 \noalign{\smallskip} 
 \hline 
 \noalign{\smallskip} 
53430	&	2005-03-01	&	D	&	18	&	5.37	&		\\
53435	&	2005-03-06	&	D	&	25	&	5.82	&		\\
53435	&	2005-03-06	&	o	&	63	&	7.28	&		\\
53442	&	2005-03-13	&	D	&	62	&	5.18	&		\\
53480	&	2005-04-20	&	D	&	63	&	8.30	&		\\
53480	&	2005-04-20	&	o	&	103	&	5.42	&		\\
53509	&	2005-05-19	&	D	&	48	&	5.62	&		\\
53559	&	2005-07-08	&	o	&	101	&	5.62	&		\\
53572	&	2005-07-21	&	D	&	34	&	11.01	&		\\
53572	&	2005-07-21	&	o	&	43	&	6.07	&		\\
53573	&	2005-07-22	&	D	&	16	&	5.72	&		\\
53612	&	2005-08-30	&	D	&	19	&	6.13	&		\\
53612	&	2005-08-30	&	o	&	71	&	6.86	&		\\
53612	&	2005-08-30	&	T	&	157 	&	7.71	&	152652	\\
53612	&	2005-08-30	&	d	&	246	&	7.73	&		\\
53639	&	2005-09-26	&	D	&	18	&	7.17	&		\\
53639	&	2005-09-26	&	o	&	47	&	5.09	&		\\
53811	&	2006-03-17	&	d	&	208	&	7.13	&		\\
53811	&	2006-03-17	&	o	&	197	&	7.47	&		\\
53875	&	2006-05-20	&	D	&	37	&	7.86	&		\\
53875	&	2006-05-20	&	o	&	96	&	5.11	&		\\
53875	&	2006-05-20	&	T	&	 21	&	5.78	&	210886	\\
53875	&	2006-05-20	&	d	&	174	&	5.33	&		\\
53898	&	2006-06-12	&	d	&	71	&	5.28	&		\\
53910	&	2006-06-24	&	d	&	375	&	5.36	&		\\
53910	&	2006-06-24	&	T	&	 191	&	5.34	&	215914	\\
53911	&	2006-06-25	&	D	&	32	&	6.70	&		\\
53911	&	2006-06-25	&	o	&	90	&	5.18	&		\\
53958	&	2006-08-11	&	D	&	22	&	5.18	&		\\
53978	&	2006-08-31	&	D	&	24	&	5.71	&		\\
54000	&	2006-09-22	&	o	&	92	&	5.18	&		\\
54015	&	2006-10-07	&	D	&	64	&	5.04	&		\\
54015	&	2006-10-07	&	o	&	74	&	5.54	&		\\
54095	&	2006-12-26	&	d	&	144	&	6.06	&		\\
54119	&	2007-01-19	&	o	&	64	&	5.74	&		\\
54167	&	2007-03-08	&	d	&	170	&	5.45	&		\\
54196	&	2007-04-06	&	d	&	94	&	6.16	&		\\
54196	&	2007-04-06	&	o	&	82	&	6.12	&		\\
54239	&	2007-05-19	&	o	&	108	&	7.33	&		\\
54239	&	2007-05-19	&	d	&	217	&	6.85	&		\\
54264	&	2007-06-13	&	o	&	70	&	6.30	&		\\
54296	&	2007-07-15	&	o	&	76	&	6.36	&		\\
54310	&	2007-07-29	&	d	&	315	&	10.00	&		\\
54310	&	2007-07-29	&	T	&	145	&	9.98	&	286412	\\
54310	&	2007-07-29	&	o	&	283	&	8.21	&		\\
54320	&	2007-08-08	&	D	&	18	&	5.04	&		\\
54320	&	2007-08-08	&	d	&	71	&	5.34	&		\\
54320	&	2007-08-08	&	o	&	93	&	5.48	&		\\
54346	&	2007-09-03	&	o	&	59	&	5.74	&		\\
54368	&	2007-09-25	&	o	&	80	&	5.66	&		\\
54368	&	2007-09-25	&	d	&	191	&	5.82	&		\\
54412	&	2007-11-08	&	D	&	22	&	5.06	&		\\
54506	&	2008-02-10	&	o	&	276	&	5.78	&		\\
54506	&	2008-02-10	&	d	&	335	&	5.64	&		\\
54525	&	2008-02-29	&	o	&	73	&	5.64	&		\\
 \noalign{\smallskip}
  \hline
   \end{tabular}
\tablefoot{
\tablefoottext{a}{Detection in the daily (D) or orbital (o) sampling; (T) BAT trigger; (d) on board detection.}
\tablefoottext{b}{Flux in mCrab the 15--50\,keV energy range.  }
\tablefoottext{c}{Double trigger. }
}  
\end{table}

\setcounter{table}{4} 
\begin{table} 	
 \tabcolsep 4pt         
 \caption{IGR~J16479$-$4514 --continued.} 	
 \begin{tabular}{lcrrrr} 
 \hline 
 \hline 
 \noalign{\smallskip} 
MJD  & Date & Detection\tablefootmark{a} & Flux\tablefootmark{b}   & S/N  &Trigger  	 	          	    \\ 
        &      &  	&     (mCrab)         & 	 &  \#    	    \\ 
 \noalign{\smallskip} 
 \hline 
 \noalign{\smallskip} 

54535	&	2008-03-10	&	D	&	20	&	6.00	&		\\
54535	&	2008-03-10	&	o	&	83	&	5.14	&		\\
54535	&	2008-03-10	&	d	&	155	&	5.24	&		\\
54544	&	2008-03-19	&	T	&	 242	&	12.02	&	{\it 306829}\tablefootmark{c} 	\\
54544	&	2008-03-19	&	d	&	1122	&	21.67	&		\\
54544	&	2008-03-19	&	o	&	1026	&	16.72	&		\\
54544	&	2008-03-19	&	T	&	 780	&	21.64	&	{\it 306830}\tablefootmark{c} 	\\
54572	&	2008-04-16	&	d	&	95	&	5.54	&		\\
54607	&	2008-05-21	&	D	&	18	&	5.58	&		\\
54607	&	2008-05-21	&	o	&	131	&	5.33	&		\\
54607	&	2008-05-21	&	d	&	217	&	7.23	&		\\
54607	&	2008-05-21	&	T	&	181	&	7.21	&	312068	\\
54664	&	2008-07-17	&	d	&	187	&	5.82	&		\\
54679	&	2008-08-01	&	o	&	226	&	8.24	&		\\
54679	&	2008-08-01	&	d	&	381	&	7.13	&		\\
54682	&	2008-08-04	&	d	&	290	&	9.30	&		\\
54687	&	2008-08-09	&	D	&	24	&	7.61	&		\\
54687	&	2008-08-09	&	d	&	83	&	5.16	&		\\
54687	&	2008-08-09	&	o	&	154	&	7.83	&		\\
54691	&	2008-08-13	&	D	&	18	&	5.31	&		\\
54826	&	2008-12-26	&	d	&	173	&	5.70	&		\\
54860	&	2009-01-29	&	D	&	26	&	7.48	&		\\
54860	&	2009-01-29	&	o	&	109	&	8.21	&		\\
54860	&	2009-01-29	&	T	&	 172	&	10.68	&	341452	\\
54860	&	2009-01-29	&	d	&	169	&	6.33	&		\\
54870	&	2009-02-08	&	o	&	89	&	5.76	&		\\
54894	&	2009-03-04	&	D	&	23	&	6.21	&		\\
54894	&	2009-03-04	&	o	&	358	&	13.22	&		\\
54894	&	2009-03-04	&	d	&	319	&	14.52	&		\\
54938	&	2009-04-17	&	D	&	26	&	5.27	&		\\
54938	&	2009-04-17	&	o	&	61	&	5.06	&		\\
54950	&	2009-04-29	&	o	&	65	&	5.09	&		\\
54957	&	2009-05-06	&	d	&	100	&	5.12	&		\\
55116	&	2009-10-12	&	d	&	96	&	5.95	&		\\
55129	&	2009-10-25	&	D	&	33	&	5.41	&		\\
55228	&	2010-02-01	&	D	&	23	&	5.79	&		\\
55228	&	2010-02-01	&	o	&	79	&	5.07	&		\\
55239	&	2010-02-12	&	D	&	18	&	5.31	&		\\
55239	&	2010-02-12	&	o	&	114	&	5.84	&		\\
55312	&	2010-04-26	&	o	&	93	&	6.01	&		\\
55342	&	2010-05-26	&	D	&	22	&	5.52	&		\\
55371	&	2010-06-24	&	o	&	101	&	7.99	&		\\
55371	&	2010-06-24	&	d	&	150	&	5.68	&		\\
55402	&	2010-07-25	&	d	&	224	&	5.69	&		\\
55420	&	2010-08-12	&	o	&	92	&	5.12	&		\\
55437	&	2010-08-29	&	o	&	101	&	6.37	&		\\
55457	&	2010-09-18	&	o	&	124	&	6.67	&		\\
55457	&	2010-09-18	&	d	&	249	&	6.76	&		\\
55513	&	2010-11-13	&	d	&	100	&	5.07	&		\\
55538	&	2010-12-08	&	o	&	96	&	5.02	&		\\
55538	&	2010-12-08	&	d	&	108	&	6.05	&		\\
55604	&	2011-02-12	&	D	&	35	&	6.92	&		\\
55604	&	2011-02-12	&	d	&	298	&	6.06	&		\\
55604	&	2011-02-12	&	o	&	179	&	11.12	&		\\
  \noalign{\smallskip}
  \hline
   \end{tabular}
\tablefoot{
\tablefoottext{a}{Detection in the daily (D) or orbital (o) sampling; (T) BAT trigger; (d) on board detection.}
\tablefoottext{b}{Flux in mCrab the 15--50\,keV energy range.  }
\tablefoottext{c}{Double trigger. }
}  
\end{table}

\setcounter{table}{4} 
\begin{table} 	
 \tabcolsep 4pt         
 \caption{IGR~J16479$-$4514 --continued.} 	
 \begin{tabular}{lcrrrr} 
 \hline 
 \hline 
  \noalign{\smallskip} 
MJD  & Date & Detection\tablefootmark{a} & Flux\tablefootmark{b}   & S/N  &Trigger  	 	          	    \\ 
        &      &  	&     (mCrab)         & 	 &  \#    	    \\ 
 \noalign{\smallskip} 
 \hline 
 \noalign{\smallskip} 
55647	&	2011-03-27	&	D	&	21	&	5.40	&		\\
55647	&	2011-03-27	&	d	&	98	&	6.62	&		\\
55647	&	2011-03-27	&	o	&	82	&	5.98	&		\\
55657	&	2011-04-06	&	D	&	38	&	6.05	&		\\
55657	&	2011-04-06	&	d	&	62	&	5.60	&		\\
55657	&	2011-04-06	&	o	&	89	&	5.20	&		\\
55663	&	2011-04-12	&	D	&	43	&	5.19	&		\\
55663	&	2011-04-12	&	o	&	85	&	5.21	&		\\
55723	&	2011-06-11	&	d	&	44	&	5.23	&		\\
55723	&	2011-06-11	&	o	&	77	&	5.51	&		\\
55739	&	2011-06-27	&	d	&	62	&	5.20	&		\\
55770	&	2011-07-28	&	o	&	103	&	5.03	&		\\
55770	&	2011-07-28	&	d	&	66	&	5.20	&		\\
55771	&	2011-07-29	&	D	&	30	&	5.18	&		\\
55805	&	2011-09-01	&	d	&	268	&	7.69	&		\\
55849	&	2011-10-15	&	o	&	94	&	8.32	&		\\
55849	&	2011-10-15	&	d	&	101	&	6.92	&		\\
55852	&	2011-10-18	&	o	&	114	&	7.17	&		\\
55852	&	2011-10-18	&	d	&	104	&	5.98	&		\\
55870	&	2011-11-05	&	d	&	140	&	5.74	&		\\
55870	&	2011-11-05	&	o	&	213	&	5.43	&		\\
56001	&	2012-03-15	&	d	&	82	&	6.19	&		\\
56001	&	2012-03-15	&	o	&	76	&	5.28	&		\\
56010	&	2012-03-24	&	o	&	79	&	5.70	&		\\
56010	&	2012-03-24	&	d	&	93	&	6.73	&		\\
56013	&	2012-03-27	&	D	&	18	&	5.89	&		\\
56013	&	2012-03-27	&	d	&	81	&	6.16	&		\\
56013	&	2012-03-27	&	o	&	83	&	5.37	&		\\
56049	&	2012-05-02	&	D	&	18	&	5.70	&		\\
56049	&	2012-05-02	&	d	&	127	&	5.10	&		\\
56049	&	2012-05-02	&	o	&	71	&	6.58	&		\\
56076	&	2012-05-29	&	o	&	131	&	7.34	&		\\
56129	&	2012-07-21	&	o	&	63	&	5.22	&		\\
56132	&	2012-07-24	&	D	&	26	&	6.05	&		\\
56179	&	2012-09-09	&	o	&	150	&	9.52	&		\\
56179	&	2012-09-09	&	d	&	242	&	9.42	&		\\
56186	&	2012-09-16	&	o	&	48	&	5.21	&		\\
56188	&	2012-09-18	&	o	&	123	&	6.10	&		\\
56196	&	2012-09-26	&	D	&	50	&	5.66	&		\\
56196	&	2012-09-26	&	d	&	203	&	6.74	&		\\
56196	&	2012-09-26	&	o	&	188	&	7.91	&		\\
56198	&	2012-09-28	&	D	&	26	&	5.37	&		\\
56229	&	2012-10-29	&	d	&	245	&	6.03	&		\\
56304	&	2013-01-12	&	D	&	47	&	5.30	&		\\
56304	&	2013-01-12	&	o	&	210	&	9.79	&		\\
56304	&	2013-01-12	&	d	&	240	&	6.86	&		\\
56341	&	2013-02-18	&	d	&	69	&	6.19	&		\\
56341	&	2013-02-18	&	o	&	94	&	5.31	&		\\
56387	&	2013-04-05	&	D	&	33	&	6.71	&		\\

  \noalign{\smallskip}
  \hline
   \end{tabular}
\tablefoot{
\tablefoottext{a}{Detection in the daily (D) or orbital (o) sampling; (T) BAT trigger; (d) on board detection.}
\tablefoottext{b}{Flux in mCrab the 15--50\,keV energy range.  }
\tablefoottext{c}{Double trigger. }
}  
\end{table}

 \setcounter{table}{5} 
 \begin{table} 	
 \tabcolsep 4pt         
 \caption{BAT detections of XTE~J1739$-$302. \label{sfxtcat:tab:17391dets}} 	
 \centering
 \begin{tabular}{rrcrrl} 
 \hline 
 \hline 
 \noalign{\smallskip} 
MJD  & Date & Detection\tablefootmark{a} & Flux\tablefootmark{b}   & S/N  &Trigger  	 	          	    \\ 
        &      &  	&     (mCrab)         & 	 &  \#    	    \\ 
 \noalign{\smallskip} 
 \hline 
 \noalign{\smallskip} 

53424	&	2005-02-23	&	D	&	23	&	6.81	&		\\
53424	&	2005-02-23	&	o	&	51	&	6.12	&		\\
53581	&	2005-07-30	&	o	&	57	&	5.73	&		\\
53581	&	2005-07-30	&	d	&	99	&	5.85	&		\\
53765	&	2006-01-30	&	D	&	47	&	7.89	&		\\
53765	&	2006-01-30	&	d	&	275	&	5.11	&		\\
53765	&	2006-01-30	&	o	&	181	&	6.52	&		\\
53798	&	2006-03-04	&	o	&	80	&	5.96	&		\\
53798	&	2006-03-04	&	d	&	130	&	7.42	&		\\
53802	&	2006-03-08	&	o	&	106	&	6.42	&		\\
53803	&	2006-03-09	&	o	&	67	&	6.32	&		\\
53806	&	2006-03-12	&	d	&	86	&	6.58	&		\\
53859	&	2006-05-04	&	o	&	88	&	5.06	&		\\
54011	&	2006-10-03	&	o	&	126	&	5.86	&		\\
54140	&	2007-02-09	&	d	&	110	&	6.61	&		\\
54140	&	2007-02-09	&	o	&	102	&	5.69	&		\\
54161	&	2007-03-02	&	d	&	250	&	7.86	&		\\
54161	&	2007-03-02	&	o	&	107	&	5.19	&		\\
54168	&	2007-03-09	&	o	&	129	&	5.83	&		\\
54168	&	2007-03-09	&	d	&	160	&	5.59	&		\\
54269	&	2007-06-18	&	d	&	183	&	5.71	&		\\
54269	&	2007-06-18	&	T	&	 40	&	6.53	&	282535	\\
54269	&	2007-06-18	&	o	&	209	&	5.41	&		\\
54347	&	2007-09-04	&	D	&	29	&	6.15	&		\\
54347	&	2007-09-04	&	o	&	165	&	8.34	&		\\
54411	&	2007-11-07	&	o	&	147	&	7.05	&		\\
54411	&	2007-11-07	&	d	&	178	&	5.89	&		\\
54564	&	2008-04-08	&	d	&	125	&	7.84	&		\\
54564	&	2008-04-08	&	T	&	 79	&	7.83	&	308797	\\
54565	&	2008-04-09	&	d	&	291	&	9.42	&		\\
54565	&	2008-04-09	&	o	&	199	&	9.06	&		\\
54632	&	2008-06-15	&	o	&	118	&	6.16	&		\\
54632	&	2008-06-15	&	d	&	190	&	5.88	&		\\
54691	&	2008-08-13	&	o	&	147	&	6.29	&		\\
54691	&	2008-08-13	&	T	&	 386	&	9.15	&	{\it 319963}\tablefootmark{c}	\\
54691	&	2008-08-13	&	d	&	160	&	5.46	&		\\
54692	&	2008-08-14	&	D	&	35	&	9.79	&		\\
54692	&	2008-08-14	&	o	&	240	&	12.72	&		\\
54692	&	2008-08-14	&	d	&	329	&	11.15	&		\\
54692	&	2008-08-14	&	T	&	 261	&	11.14	&	{\it 319964}\tablefootmark{c}	\\
54724	&	2008-09-15	&	d	&	188	&	5.01	&		\\
54900	&	2009-03-10	&	o	&	93	&	7.90	&		\\
54900	&	2009-03-10	&	d	&	105	&	6.83	&		\\
54900	&	2009-03-10	&	T	&	 50	&	6.81	&	346069	\\
55079	&	2009-09-05	&	d	&	91	&	7.20	&		\\
55107	&	2009-10-03	&	d	&	181	&	6.32	&		\\
55202	&	2010-01-06	&	o	&	153	&	5.14	&		\\
55226	&	2010-01-30	&	d	&	166	&	7.50	&		\\
55226	&	2010-01-30	&	o	&	139	&	5.03	&		\\
55227	&	2010-01-31	&	o	&	128	&	6.91	&		\\
55227	&	2010-01-31	&	d	&	167	&	8.69	&		\\
55276	&	2010-03-21	&	o	&	104	&	5.31	&		\\
55276	&	2010-03-21	&	d	&	116	&	6.50	&		\\
55408	&	2010-07-31	&	d	&	190	&	6.38	&		\\
55408	&	2010-07-31	&	o	&	166	&	7.59	&		\\
 \noalign{\smallskip}
  \hline
   \end{tabular}
\tablefoot{
\tablefoottext{a}{Detection in the daily (D) or orbital (o) sampling; (T) BAT trigger; (d) on board detection.}
\tablefoottext{b}{Flux in mCrab the 15--50\,keV energy range.  }
\tablefoottext{c}{Double trigger. }
}  
\end{table}

\setcounter{table}{5} 
 \begin{table} 	
 \tabcolsep 4pt         
 \caption{XTE~J1739$-$302--continued.} 	
 \centering
 \begin{tabular}{rrcrrl} 
 \hline 
 \hline 
 \noalign{\smallskip} 
MJD  & Date & Detection\tablefootmark{a} & Flux\tablefootmark{b}   & S/N  &Trigger  	 	          	    \\ 
        &      &  	&     (mCrab)         & 	 &  \#    	    \\ 
 \noalign{\smallskip} 
 \hline 
 \noalign{\smallskip} 
55432	&	2010-08-24	&	d	&	85	&	6.02	&		\\
55573	&	2011-01-12	&	d	&	93	&	6.15	&		\\
55585	&	2011-01-24	&	o	&	106	&	5.35	&		\\
55585	&	2011-01-24	&	d	&	180	&	5.51	&		\\
55609	&	2011-02-17	&	d	&	170	&	5.84	&		\\
55614	&	2011-02-22	&	d	&	288	&	7.47	&		\\
55614	&	2011-02-22	&	T	&	 309	&	7.44	&	446475	\\
55614	&	2011-02-22	&	o	&	260	&	5.98	&		\\
55622	&	2011-03-02	&	d	&	364	&	12.10	&		\\
55622	&	2011-03-02	&	o	&	415	&	10.22	&		\\
55708	&	2011-05-27	&	d	&	68	&	5.46	&		\\
55708	&	2011-05-27	&	o	&	116	&	5.12	&		\\
55795	&	2011-08-22	&	d	&	196	&	5.95	&		\\
55879	&	2011-11-14	&	d	&	136	&	5.18	&		\\
55985	&	2012-02-28	&	d	&	205	&	5.95	&		\\
55986	&	2012-02-29	&	d	&	162	&	5.45	&		\\
56179	&	2012-09-09	&	d	&	84	&	5.82	&		\\
56179	&	2012-09-09	&	T	&	 62	&	5.81	&	533120	\\
56242	&	2012-11-11	&	D	&	31	&	5.11	&		\\
56242	&	2012-11-11	&	T	&	 119	&	6.10	&	538084	\\
56242	&	2012-11-11	&	d	&	147	&	5.26	&		\\
56242	&	2012-11-11	&	o	&	150	&	6.30	&		\\
56326	&	2013-02-03	&	d	&	167	&	6.11	&		\\
56379	&	2013-03-28	&	d	&	180	&	7.01	&		\\

   \noalign{\smallskip}
  \hline
   \end{tabular}
\tablefoot{
\tablefoottext{a}{Detection in the daily (D) or orbital (o) sampling; (T) BAT trigger; (d) on board detection.}
\tablefoottext{b}{Flux in mCrab the 15--50\,keV energy range.  }
\tablefoottext{c}{Double trigger. }
}  
\end{table} 

 \setcounter{table}{6} 
 \begin{table} 	
 \tabcolsep 4pt         
 \caption{BAT detections of IGR~J17544$-$2619.  \label{sfxtcat:tab:17544dets}} 	
 \centering
 \begin{tabular}{rrcrrl} 
 \hline 
 \hline 
 \noalign{\smallskip} 
MJD  & Date & Detection\tablefootmark{a} & Flux\tablefootmark{b}   & S/N  &Trigger  	 	          	    \\ 
        &      &  	&     (mCrab)         & 	 &  \#    	    \\ 
 \noalign{\smallskip} 
 \hline 
 \noalign{\smallskip} 
53758	&	2006-01-23	&	D	&	71	&	6.74	&		\\
53758	&	2006-01-23	&	o	&	71	&	7.30	&		\\
53811	&	2006-03-17	&	D	&	27	&	6.36	&		\\
53998	&	2006-09-20	&	o	&	86	&	5.80	&		\\
54035	&	2006-10-27	&	d	&	97	&	5.20	&		\\
54372	&	2007-09-29	&	D	&	24	&	6.16	&		\\
54372	&	2007-09-29	&	o	&	113	&	8.34	&		\\
54372	&	2007-09-29	&	d	&	220	&	7.15	&		\\
54387	&	2007-10-14	&	D	&	19	&	7.50	&		\\
54387	&	2007-10-14	&	o	&	148	&	8.71	&		\\
54387	&	2007-10-14	&	d	&	221	&	7.61	&		\\
54388	&	2007-10-15	&	D	&	28	&	10.39	&		\\
54388	&	2007-10-15	&	o	&	77	&	5.32	&		\\
54412	&	2007-11-08	&	D	&	22	&	6.99	&		\\
54412	&	2007-11-08	&	o	&	167	&	6.22	&		\\
54412	&	2007-11-08	&	d	&	215	&	7.75	&		\\
54555	&	2008-03-30	&	o	&	74	&	5.29	&		\\
54556   & 2008-03-31 &o\tablefootmark{c} & 51	& 4.87 \\  
54556	&	2008-03-31	&	d	&	200	&	9.10	&		\\
54556	&	2008-03-31	&	T	&	 	&	9.10	&	308224	\\
54565	&	2008-04-09	&	D	&	31	&	6.11	&		\\
54565	&	2008-04-09	&	d	&	76	&	5.29	&		\\
54565	&	2008-04-09	&	o	&	70	&	5.31	&		\\
54611	&	2008-05-25	&	D	&	62	&	6.59	&		\\
54611	&	2008-05-25	&	o	&	308	&	9.35	&		\\
54611	&	2008-05-25	&	d	&	230	&	6.51	&		\\
54708	&	2008-08-30	&	D	&	21	&	5.54	&		\\
54708	&	2008-08-30	&	o	&	54	&	5.11	&		\\
54988	&	2009-06-06	&	o	&	170	&	8.41	&		\\
54988	&	2009-06-06	&	d	&	235	&	8.16	&		\\
54988	&	2009-06-06	&	T	&	 	&	8.15	&	354221	\\
55024	&	2009-07-12	&	d	&	90	&	5.29	&		\\
55063	&	2009-08-20	&	o	&	116	&	6.09	&		\\
55063	&	2009-08-20	&	d	&	148	&	5.54	&		\\
55087	&	2009-09-13	&	o	&	131	&	5.86	&		\\
55259	&	2010-03-04	&	o	&	88	&	6.10	&		\\
55259	&	2010-03-04	&	d	&	131	&	7.47	&		\\
55259	&	2010-03-04	&	T	&	 87	&	7.45	&	414875	\\
55260	&	2010-03-05	&	D	&	24	&	6.29	&		\\
55260	&	2010-03-05	&	o	&	144	&	7.73	&		\\
55260	&	2010-03-05	&	d	&	223	&	7.20	&		\\
55364	&	2010-06-17	&	d	&	91	&	5.52	&		\\
55610	&	2011-02-18	&	o	&	74	&	6.19	&		\\
55610	&	2011-02-18	&	d	&	182	&	6.20	&		\\
55644	&	2011-03-24	&	D	&	35	&	6.50	&		\\
55644	&	2011-03-24	&	d	&	395	&	12.82	&		\\
55644	&	2011-03-24	&	T	&	 223	&	12.78	&	449907	\\
55644	&	2011-03-24	&	o	&	94	&	9.42	&		\\
55668	&	2011-04-17	&	o	&	178	&	6.81	&		\\
55668	&	2011-04-17	&	d	&	198	&	6.59	&		\\
55748	&	2011-07-06	&	d	&	77	&	5.42	&		\\
55999	&	2012-03-13	&	o	&	151	&	5.77	&		\\
55999	&	2012-03-13	&	d	&	242	&	5.81	&		\\

  \noalign{\smallskip}
  \hline
   \end{tabular}
\tablefoot{
\tablefoottext{a}{Detection in the daily (D) or orbital (o) sampling; (T) BAT trigger; (d) on board detection.}
\tablefoottext{b}{Flux in mCrab the 15--50\,keV energy range.  }
\tablefoottext{c}{Short (112s) flare that triggered onboard.  The BATTM monitor result here reported 
is for a 688s interval, at a lower total significance.}
}  
\end{table} 

\setcounter{table}{6} 
 \begin{table} 	
 \tabcolsep 4pt         
 \caption{IGR~J17544$-$2619--continued.  } 	
  \centering
 \begin{tabular}{rrcrrl} 
 \hline 
 \hline 
 \noalign{\smallskip} 
MJD  & Date & Detection\tablefootmark{a} & Flux\tablefootmark{b}   & S/N  &Trigger  	 	          	    \\ 
        &      &  	&     (mCrab)         & 	 &  \#    	    \\ 
 \noalign{\smallskip} 
 \hline 
 \noalign{\smallskip} 
56004	&	2012-03-18	&	d	&	150	&	5.37	&		\\
56029	&	2012-04-12	&	d	&	104	&	5.69	&		\\
56029	&	2012-04-12	&	o	&	92	&	5.08	&		\\
56061	&	2012-05-14	&	o	&	82	&	5.26	&		\\
56132	&	2012-07-24	&	D	&	23	&	6.34	&		\\
56132	&	2012-07-24	&	d	&	93	&	6.14	&		\\
56132	&	2012-07-24	&	T	&	 40	&	6.14	&	528432	\\
56132	&	2012-07-24	&	o	&	90	&	5.11	&		\\
56240	&	2012-11-09	&	o	&	291	&	7.71	&		\\
56240	&	2012-11-09	&	d	&	216	&	7.67	&		\\
   
  \noalign{\smallskip}
  \hline
   \end{tabular}
\tablefoot{
\tablefoottext{a}{Detection in the daily (D) or orbital (o) sampling; (T) BAT trigger; (d) on board detection.}
\tablefoottext{b}{Flux in mCrab the 15--50\,keV energy range.  }
\tablefoottext{c}{Double trigger. }
\tablefoottext{d}{Short (64s) flare that triggered onboard.  The BATTM monitor result here reported 
is for a 1000s interval, at a lower total significance.}
}  
\end{table}

 \setcounter{table}{7} 
 \begin{table} 	
 \tabcolsep 4pt         
 \caption{BAT detections of SAX~J1818.6$-$1703. \label{sfxtcat:tab:18186dets}} 	
 \centering
 \begin{tabular}{rrcrrl} 
 \hline 
 \hline 
 \noalign{\smallskip} 
MJD  & Date & Detection\tablefootmark{a} & Flux\tablefootmark{b}   & S/N  &Trigger  	 	          	    \\ 
        &      &  	&     (mCrab)         & 	 &  \#    	    \\ 
 \noalign{\smallskip} 
 \hline 
 \noalign{\smallskip} 
53850	&	2006-04-25	&	D	&	26	&	5.11	&		\\
53850	&	2006-04-25	&	o	&	65	&	5.74	&		\\
53970	&	2006-08-23	&	d	&	105	&	6.32	&		\\
54000	&	2006-09-22	&	o	&	89	&	5.66	&		\\
54000	&	2006-09-22	&	d	&	218	&	5.88	&		\\
54360	&	2007-09-17	&	D	&	23	&	6.69	&		\\
54360	&	2007-09-17	&	d	&	121	&	6.73	&		\\
54360	&	2007-09-17	&	o	&	133	&	6.79	&		\\
54389	&	2007-10-16	&	D	&	28	&	8.11	&		\\
54389	&	2007-10-16	&	o	&	101	&	9.69	&		\\
54389	&	2007-10-16	&	T	&	132	&	7.94	&	294385	\\
54389	&	2007-10-16	&	d	&	171	&	5.98	&		\\
54510	&	2008-02-14	&	d	&	96	&	5.17	&		\\
54510	&	2008-02-14	&	o	&	139	&	6.27	&		\\
54540	&	2008-03-15	&	T	&	 57	&	8.02	&	306379	\\
54540	&	2008-03-15	&	o	&	133	&	6.49	&		\\
54540	&	2008-03-15	&	d	&	185	&	5.58	&		\\
54541	&	2008-03-16	&	o	&	61	&	5.96	&		\\
54572	&	2008-04-16	&	o	&	105	&	5.87	&		\\
54632	&	2008-06-15	&	D	&	35	&	5.72	&		\\
54632	&	2008-06-15	&	o	&	137	&	6.24	&		\\
54632	&	2008-06-15	&	d	&	124	&	6.91	&		\\
54957	&	2009-05-06	&	D	&	35	&	9.79	&		\\
54957	&	2009-05-06	&	o	&	231	&	8.45	&		\\
54957	&	2009-05-06	&	d	&	228	&	7.15	&		\\
54957	&	2009-05-06	&	T	&	149 	&	7.12	&	351323	\\
55079	&	2009-09-05	&	D	&	46	&	11.01	&		\\
55079	&	2009-09-05	&	T	&	87 	&	6.63	&	361958	\\
55079	&	2009-09-05	&	o	&	139	&	13.15	&		\\
55079	&	2009-09-05	&	d	&	159	&	6.21	&		\\
55139	&	2009-11-04	&	D	&	58	&	10.14	&		\\
55139	&	2009-11-04	&	T	&	100	&	7.34	&	374869	\\
55139	&	2009-11-04	&	o	&	210	&	7.26	&		\\
55139	&	2009-11-04	&	d	&	224	&	6.55	&		\\
55171	&	2009-12-06	&	d	&	77	&	5.95	&		\\
55440	&	2010-09-01	&	o	&	99	&	5.40	&		\\
55595	&	2011-02-03	&	d	&	181	&	5.59	&		\\
55801	&	2011-08-28	&	o	&	91	&	5.01	&		\\
55802	&	2011-08-29	&	D	&	27	&	5.32	&		\\
56009	&	2012-03-23	&	o	&	65	&	5.13	&		\\
56098	&	2012-06-20	&	o	&	84	&	5.50	&		\\
56130	&	2012-07-22	&	d	&	73	&	5.67	&		\\
56368	&	2013-03-17	&	d	&	146	&	5.74	&		\\
56368	&	2013-03-17	&	o	&	104	&	6.07	&		\\

 \noalign{\smallskip}
  \hline
   \end{tabular}
\tablefoot{
\tablefoottext{a}{Detection in the daily (D) or orbital (o) sampling; (T) BAT trigger; (d) on board detection.}
\tablefoottext{b}{Flux in mCrab the 15--50\,keV energy range.  }
}  
\end{table} 

\setcounter{table}{8} 
 \begin{table} 	
 \tabcolsep 4pt         
 \caption{BAT detections of AX~J1841.0$-$0536. \label{sfxtcat:tab:18410dets}} 	
 \centering
 \begin{tabular}{rrcrrl} 
 \hline 
 \hline 
 \noalign{\smallskip} 
MJD  & Date & Detection\tablefootmark{a} & Flux\tablefootmark{b}   & S/N  &Trigger  	 	          	    \\ 
        &      &  	&     (mCrab)         & 	 &  \#    	    \\ 
 \noalign{\smallskip} 
 \hline 
 \noalign{\smallskip} 
53518	&	2005-05-28	&	o	&	54	&	5.11	&		\\
53530	&	2005-06-09	&	D	&	15	&	5.49	&		\\
53542	&	2005-06-21	&	o	&	74	&	5.56	&		\\
53821	&	2006-03-27	&	D	&	33	&	11.58	&		\\
53821	&	2006-03-27	&	o	&	209	&	13.51	&		\\
53821	&	2006-03-27	&	d	&	232	&	8.91	&		\\
53845	&	2006-04-20	&	o	&	77	&	5.47	&		\\
54139	&	2007-02-08	&	d	&	147	&	6.13	&		\\
54139	&	2007-02-08	&	o	&	129	&	6.80	&		\\
54195	&	2007-04-05	&	d	&	124	&	6.44	&		\\
55352	&	2010-06-05	&	D	&	25	&	5.68	&		\\
55352	&	2010-06-05	&	T	&	 46	&	6.87	&	423958	\\
55352	&	2010-06-05	&	d	&	103	&	6.21	&		\\
55352	&	2010-06-05	&	o	&	83	&	6.90	&		\\
55507	&	2010-11-07	&	d	&	78	&	5.35	&		\\
55526	&	2010-11-26	&	d	&	61	&	5.43	&		\\
55526	&	2010-11-26	&	o	&	81	&	5.32	&		\\
55736	&     2011-06-24	&      o     &     130	&     5.80 &   	        \\
55736	&	2011-06-24	&	T	&	 66	&	6.91	&	455967	\\
55736	&	2011-06-24	&	d	&    106	&	5.75	&		\\
55795	&	2011-08-22	&	d	&	89	&	6.14	&		\\
55852	&	2011-10-18	&	D	&	22	&	5.49	&		\\
55853	&	2011-10-19	&	D	&	19	&	6.50	&		\\
55853	&	2011-10-19	&	d	&	75	&	5.19	&		\\
55853	&	2011-10-19	&	o	&	60	&	5.04	&		\\
55977	&	2012-02-20	&	d	&	66	&	5.26	&		\\
55987	&	2012-03-01	&	o	&	51	&	5.28	&		\\
56092	&	2012-06-14	&	D	&	37	&	7.27	&		\\
56092	&	2012-06-14	&	T	&	 159	&	8.47	&	524364	\\
56092	&	2012-06-14	&	o	&	216	&	8.54	&		\\
56092	&	2012-06-14	&	d	&	183	&	7.99	&		\\
56099	&	2012-06-21	&	d	&	58	&	5.09	&		\\
56128	&	2012-07-20	&	d	&	58	&	5.00	&		\\
56129	&	2012-07-21	&	o	&	95	&	5.02	&		\\
56132	&	2012-07-24	&	D	&	28	&	6.41	&		\\
56132	&	2012-07-24	&	T	&	 111	&	9.81	&	528411	\\
56132	&	2012-07-24	&	o	&	198	&	6.92	&		\\
56132	&	2012-07-24	&	d	&	175	&	5.84	&		\\
56150	&	2012-08-11	&	o	&	79	&	5.86	&		\\
56201	&	2012-10-01	&	D	&	23	&	5.54	&		\\
56238	&	2012-11-07	&	d	&	156	&	6.29	&		\\
56353	&	2013-03-02	&	o	&	240	&	11.81	&		\\
56353	&	2013-03-02	&	d	&	275	&	11.06	&		\\
56354	&	2013-03-03	&	d	&	87	&	5.42	&		\\
56354	&	2013-03-03	&	o	&	163	&	5.59	&		\\

 \noalign{\smallskip}
  \hline
   \end{tabular}
\tablefoot{
\tablefoottext{a}{Detection in the daily (D) or orbital (o) sampling; (T) BAT trigger; (d) on board detection.}
\tablefoottext{b}{Flux in mCrab the 15--50\,keV energy range.  }
}  
\end{table} 

\setcounter{table}{9} 
 \begin{table} 	
 \tabcolsep 4pt         
 \caption{BAT detections of AX~J1845.0$-$0433. \label{sfxtcat:tab:18450dets}} 	
 \centering
 \begin{tabular}{rrcrrl} 
 \hline 
 \hline 
 \noalign{\smallskip} 
MJD  & Date & Detection\tablefootmark{a} & Flux\tablefootmark{b}   & S/N  &Trigger  	 	          	    \\ 
        &      &  	&     (mCrab)         & 	 &  \#    	    \\ 
 \noalign{\smallskip} 
 \hline 
 \noalign{\smallskip} 

53536	&	2005-06-15	&	D	&	22	&	6.03	&		\\
53678	&	2005-11-04	&	o	&	77	&	5.65	&		\\
53678	&	2005-11-04	&	d	&	83	&	6.40	&		\\
53678	&	2005-11-04	&	T	&	 33	&	5.43	&	162526	\\
53786	&	2006-02-20	&	o	&	135	&	5.61	&		\\
53859	&	2006-05-04	&	o	&	78	&	5.80	&		\\
54004	&	2006-09-26	&	d	&	101	&	5.54	&		\\
54004	&	2006-09-26	&	o	&	136	&	5.66	&		\\
54282	&	2007-07-01	&	d	&	128	&	5.20	&		\\
54540	&	2008-03-15	&	d	&	100	&	6.79	&		\\
54540	&	2008-03-15	&	o	&	97	&	5.56	&		\\
54697	&	2008-08-19	&	o	&	117	&	5.47	&		\\
54782	&	2008-11-12	&	d	&	76	&	5.53	&		\\
54967	&	2009-05-16	&	d	&	60	&	5.35	&		\\
55010	&	2009-06-28	&	d	&	80	&	6.12	&		\\
55010	&	2009-06-28	&	o	&	87	&	6.05	&		\\
55010	&	2009-06-28	&	T	&	 76	&	6.99	&	355911	\\
55947	&	2012-01-21	&	D	&	32	&	5.08	&		\\
56052	&	2012-05-05	&	D	&	19	&	5.38	&		\\
56052	&	2012-05-05	&	o	&	220	&	5.36	&		\\
56052	&	2012-05-05	&	T	&	 134	&	7.26	&	521567	\\
56052	&	2012-05-05	&	d	&	148	&	6.01	&		\\
56161	&	2012-08-22	&	d	&	160	&	5.36	&		\\

 \noalign{\smallskip}
  \hline
   \end{tabular}
\tablefoot{
\tablefoottext{a}{Detection in the daily (D) or orbital (o) sampling; (T) BAT trigger; (d) on board detection.}
\tablefoottext{b}{Flux in mCrab the 15--50\,keV energy range.  }
}  
\end{table} 
 \setcounter{table}{10} 
 \begin{table} 	
 \tabcolsep 4pt         
 \caption{BAT detections of IGR~J18483$-$0311. \label{sfxtcat:tab:18483dets}} 	
 \centering
 \begin{tabular}{rrcrrl} 
 \hline 
 \hline 
 \noalign{\smallskip} 
MJD  & Date & Detection\tablefootmark{a} & Flux\tablefootmark{b}   & S/N  &Trigger  	 	          	    \\ 
        &      &  	&     (mCrab)         & 	 &  \#    	    \\ 
 \noalign{\smallskip} 
 \hline 
 \noalign{\smallskip} 
53439	&	2005-03-10	&	D	&	16	&	5.61	&		\\
53445	&	2005-03-16	&	D	&	19	&	5.30	&		\\
53643	&	2005-09-30	&	o	&	69	&	5.05	&		\\
53775	&	2006-02-09	&	d	&	169	&	5.61	&		\\
53844	&	2006-04-19	&	D	&	24	&	7.24	&		\\
53844	&	2006-04-19	&	o	&	56	&	5.43	&		\\
53844	&	2006-04-19	&	d	&	124	&	5.80	&		\\
53845	&	2006-04-20	&	D	&	27	&	5.47	&		\\
53901	&	2006-06-15	&	D	&	38	&	7.49	&		\\
53901	&	2006-06-15	&	o	&	71	&	5.22	&		\\
53901	&	2006-06-15	&	d	&	88	&	5.02	&		\\
53959	&	2006-08-12	&	D	&	22	&	5.02	&		\\
53961	&	2006-08-14	&	D	&	21	&	5.28	&		\\
53994	&	2006-09-16	&	D	&	51	&	6.57	&		\\
53994	&	2006-09-16	&	d	&	151	&	5.26	&		\\
54163	&	2007-03-04	&	D	&	35	&	7.66	&		\\
54255	&	2007-06-04	&	d	&	63	&	5.07	&		\\
54272	&	2007-06-21	&	D	&	27	&	5.50	&		\\
54364	&	2007-09-21	&	D	&	36	&	11.79	&		\\
54364	&	2007-09-21	&	d	&	207	&	6.78	&		\\
54364	&	2007-09-21	&	o	&	149	&	5.36	&		\\
54367	&	2007-09-24	&	D	&	23	&	6.22	&		\\
54367	&	2007-09-24	&	o	&	84	&	6.18	&		\\
54367	&	2007-09-24	&	d	&	77	&	5.20	&		\\
54512	&	2008-02-16	&	d	&	234	&	7.00	&		\\
54532	&	2008-03-07	&	d	&	169	&	5.32	&		\\
54552	&	2008-03-27	&	d	&	66	&	5.60	&		\\
54552	&	2008-03-27	&	o	&	62	&	5.02	&		\\
54626	&	2008-06-09	&	d	&	104	&	5.58	&		\\
54626	&	2008-06-09	&	o	&	135	&	6.69	&		\\
54702	&	2008-08-24	&	D	&	54	&	11.79	&		\\
54702	&	2008-08-24	&	d	&	265	&	11.50	&		\\
54702	&	2008-08-24	&	T	&	 235	&	11.48	&	321750	\\
54702	&	2008-08-24	&	o	&	131	&	11.20	&		\\
54903	&	2009-03-13	&	D	&	33	&	7.24	&		\\
54903	&	2009-03-13	&	d	&	76	&	6.06	&		\\
54922	&	2009-04-01	&	d	&	93	&	6.07	&		\\
54925	&	2009-04-04	&	D	&	27	&	5.17	&		\\
55014	&	2009-07-02	&	D	&	48	&	6.84	&		\\
55108	&	2009-10-04	&	D	&	46	&	8.19	&		\\
55108	&	2009-10-04	&	o	&	69	&	5.83	&		\\
55108	&	2009-10-04	&	d	&	75	&	5.29	&		\\
55109	&	2009-10-05	&	D	&	33	&	5.91	&		\\
55114	&	2009-10-10	&	D	&	31	&	6.68	&		\\
55114	&	2009-10-10	&	o	&	140	&	5.48	&		\\
55277	&	2010-03-22	&	D	&	30	&	7.37	&		\\
55277	&	2010-03-22	&	o	&	84	&	5.71	&		\\
55277	&	2010-03-22	&	d	&	96	&	5.16	&		\\
55301	&	2010-04-15	&	D	&	36	&	8.90	&		\\
55301	&	2010-04-15	&	o	&	84	&	7.21	&		\\
55301	&	2010-04-15	&	d	&	156	&	5.74	&		\\
55302	&	2010-04-16	&	D	&	28	&	5.75	&		\\ \noalign{\smallskip}
55311	&	2010-04-25	&	D	&	29	&	8.32	&		\\
55311	&	2010-04-25	&	o	&	97	&	5.08	&		\\
55312	&	2010-04-26	&	D	&	23	&	6.00	&		\\
 \hline
   \end{tabular}
\tablefoot{
\tablefoottext{a}{Detection in the daily (D) or orbital (o) sampling; (T) BAT trigger; (d) on board detection.}
\tablefoottext{b}{Flux in mCrab the 15--50\,keV energy range.  }
}  
\end{table} 

\setcounter{table}{10} 
\begin{table} 	
 \tabcolsep 4pt         
 \caption{IGR~J18483$-$0311--continued.} 	
  \centering
 \begin{tabular}{rrcrrl} 
 \hline 
 \hline 
 \noalign{\smallskip} 
MJD  & Date & Detection\tablefootmark{a} & Flux\tablefootmark{b}   & S/N  &Trigger  	 	          	    \\ 
        &      &  	&     (mCrab)         & 	 &  \#    	    \\ 
 \noalign{\smallskip} 
 \hline 
 \noalign{\smallskip} 
55366	&	2010-06-19	&	D	&	40	&	6.32	&		\\
55366	&	2010-06-19	&	o	&	145	&	7.68	&		\\
55366	&	2010-06-19	&	d	&	129	&	5.40	&		\\
55496	&	2010-10-27	&	d	&	50	&	5.54	&		\\
55497	&	2010-10-28	&	d	&	81	&	6.01	&		\\
55516	&	2010-11-16	&	o	&	91	&	6.93	&		\\
55516	&	2010-11-16	&	d	&	81	&	5.45	&		\\
55517	&	2010-11-17	&	D	&	29	&	5.90	&		\\
55517	&	2010-11-17	&	d	&	73	&	6.13	&		\\
55517	&	2010-11-17	&	o	&	62	&	5.70	&		\\
55607	&	2011-02-15	&	D	&	40	&	7.56	&		\\
55607	&	2011-02-15	&	d	&	82	&	6.18	&		\\
55625	&	2011-03-05	&	D	&	33	&	6.86	&		\\
55627	&	2011-03-07	&	d	&	80	&	5.89	&		\\
55792	&	2011-08-19	&	d	&	86	&	5.05	&		\\
55792	&	2011-08-19	&	o	&	113	&	5.79	&		\\
55815	&	2011-09-11	&	o	&	69	&	5.11	&		\\
55830	&	2011-09-26	&	D	&	24	&	5.18	&		\\
55870	&	2011-11-05	&	D	&	33	&	7.95	&		\\
55885	&	2011-11-20	&	d	&	72	&	6.34	&		\\
55888	&	2011-11-23	&	D	&	48	&	7.09	&		\\
55888	&	2011-11-23	&	d	&	58	&	5.02	&		\\
56002	&	2012-03-16	&	D	&	27	&	5.39	&		\\
56005	&	2012-03-19	&	D	&	23	&	5.16	&		\\
56006	&	2012-03-20	&	D	&	23	&	5.10	&		\\
56013	&	2012-03-27	&	D	&	22	&	5.11	&		\\
56013	&	2012-03-27	&	o	&	49	&	5.01	&		\\
56017	&	2012-03-31	&	D	&	29	&	6.07	&		\\
56017	&	2012-03-31	&	d	&	94	&	7.28	&		\\
56017	&	2012-03-31	&	o	&	82	&	5.66	&		\\
56148	&	2012-08-09	&	D	&	57	&	9.45	&		\\
56148	&	2012-08-09	&	d	&	146	&	5.26	&		\\
56148	&	2012-08-09	&	o	&	118	&	6.13	&		\\
56163	&	2012-08-24	&	D	&	31	&	7.43	&		\\
56163	&	2012-08-24	&	d	&	82	&	5.71	&		\\
56163	&	2012-08-24	&	o	&	116	&	5.14	&		\\
56164	&	2012-08-25	&	d	&	94	&	7.39	&		\\
56165	&	2012-08-26	&	o	&	79	&	6.76	&		\\
56165	&	2012-08-26	&	d	&	127	&	5.00	&		\\
56333	&	2013-02-10	&	D	&	48	&	7.80	&		\\
56333	&	2013-02-10	&	o	&	81	&	5.59	&		\\
56333	&	2013-02-10	&	d	&	99	&	7.60	&		\\
56334	&	2013-02-11	&	D	&	32	&	6.30	&		\\
56334	&	2013-02-11	&	d	&	92	&	6.85	&		\\
56443	&	2013-05-31	&	D	&	44	&	7.41	&		\\
56443	&	2013-05-31	&	d	&	64	&	5.23	&		\\ \noalign{\smallskip}
  \hline
   \end{tabular}
\tablefoot{
\tablefoottext{a}{Detection in the daily (D) or orbital (o) sampling; (T) BAT trigger; (d) on board detection.}
\tablefoottext{b}{Flux in mCrab the 15--50\,keV energy range.  }
}  
\end{table} 

\end{appendix}

\end{document}